\documentclass[preprint,nofootinbib,eqsecnum,12pt]{revtex4}
\usepackage{graphicx}
\usepackage{axodraw}

\newcount\revtex\revtex=1  %%% (1 revtex4, 0 proc)
\ifnum\revtex=1

\def\pcmatrix{\pmatrix}
\def\myinst{\affiliation}
\else
\def\pcmatrix#1{\begin{pmatrix} #1 \end{pmatrix}}
\def\myinst{\institute}
\fi

\newcount\ans\ans=0  %%% (1 answers, 0 no answers)
\ifnum\ans=0
\def\Answer#1{}
\else
\def\Answer#1{\\ \noindent
{\underline{Answer:} #1} \vspace*{3mm}}

\fi

\newcounter{quecount}
\def\que#1{\noindent{\bf  Question
\arabic{quecount}}{: #1}\addtocounter{quecount}{1} \vspace*{2mm}}

\newcommand{\ov}{\overline}
\def\vev#1{\langle #1 \rangle}
\renewcommand{\Im}{\mbox{{\cal I}m}} 
\renewcommand{\Re}{\mbox{{\cal R}e}}  

\newcommand{\beq}{\begin{equation}} 
\newcommand{\eeq}{\end{equation}}
\newcommand{\beqa}{\begin{eqnarray}}
\newcommand{\eeqa}{\end{eqnarray}}

\newcommand{\bei}{\begin{itemize}}
\newcommand{\eei}{\end{itemize}}
\newcommand{\ben}{\begin{enumerate}}
\newcommand{\een}{\end{enumerate}}

\newcommand{\lsim}{\lesssim}
\newcommand{\re}[1]{\ensuremath{{\cal R}e(#1)}}
\newcommand{\im}[1]{\ensuremath{{\cal I}m(#1)}}

\newcommand{\M}{P}
\newcommand{\Mb}{\overline{P}}
\newcommand{\Bb}{\overline{B}}
\newcommand{\fb}{\overline{f}}

\newcommand{\CP}{CP\ }
\newcommand{\Mz}{P{}^0}
\newcommand{\Mzb}{\overline{P}{}^0}
\newcommand{\Heff}{{\cal H}}
\newcommand{\Meff}{M}
\newcommand{\Geff}{\Gamma}
\newcommand{\no}{\nonumber}

\begin{document}

\title{Introduction to flavor physics\ifnum\revtex=1\footnote{%
Lectures given at the ``2009 European School of High-Energy Physics,''
Bautzen, Germany, June 14-27, 2009, and at the ``Flavianet School on
Flavor Physics,'' Karlsruhe, Germany, September 7-18, 2009.}\fi}
\author{Yuval Grossman}
\ifnum\revtex=1
\email{yg73@cornell.edu}
\fi

\myinst{
\vspace*{4mm}Institute for High Energy Phenomenology\\
Newman Laboratory of Elementary Particle Physics\\ 
Cornell University,~Ithaca,~NY~14853,~USA
\vspace*{6mm}}

\ifnum\revtex=0
\maketitle
\fi
%%%%%%%%%%%%%%%%%%%%%%%%%%%%%%%%
%\date{\today}
%\pacs{}

\ifnum\revtex=1
\vspace{1cm}
\fi
\begin{abstract}

This set of lectures covers the very basics of flavor physics and 
are aimed to be an entry point to the subject. A lot of problems are
provided in the hope of making the manuscript a self study guide.

\end{abstract}

\ifnum\revtex=1
\maketitle
\fi
%%%%%%%%%%%%%%%%%%%%%%%%%%%%%%%%%%%%%%%%%%%%%%%%%%%%%%%%%%%%%%%%%%%%%%%%%%

\section{Welcome statement}

My plan for these lectures is to introduce you to the very basics of
flavor physics. Hopefully, after the lectures you will have enough
knowledge and more importantly, enough curiosity, that you will go on
and learn more about the subject.

These are lecture notes and are not meant to be a review. I try to
present the basic ideas, hoping to give a clear
picture of the physics. Thus, many details are omitted, implicit
assumptions are made, and no references are given. Yet details are
important: after you go over the current lecture notes once or twice,
I hope you will feel the need for more.  Then it will be the time to
turn to the many reviews~\cite{Gedalia:2010rj,Isidori:2010gz,Buras:2009if,Nierste:2009wg,Artuso:2009jw,Nir:2007xn,Hocker:2006xb,Neubert:2005mu,Nir:2005js,Buras:2005xt,Ligeti:2003fi} and books~\cite{Bigi:2000yz,Branco:1999fs} on the subject.

I have tried to include many homework problems for the reader to solve, much
more than what I gave in the actual lectures. If you would like to
learn the material, I think that the provided problems are the way to
start. They force you to fully understand the issues and apply your
knowledge to new situations. The problems are given at the end of each
section.
%At the
%end of the notes I give short answers, but not the full details of the
%answers. First try to solve the homework without looking at at the
%answers. It is much nicer to solve it alone, and you will gain a lot
%of physics by doing that. 
The questions can be challenging and may take a lot of time. Do not
give up after a few minutes!

%%%%%%%%%%%%%%%%%%%%%%%%
\section{The standard model: a reminder}

I assume that you have basic knowledge of Quantum Field Theory (QFT)
and that you are familiar with the Standard Model (SM). Nevertheless, I
start with a brief review of the SM, not only to remind you, but
also since I like to present things in a way that may be different from
the way you got to know the SM.

%%%%%%%%%%%%%%%%%%%%%%%%%%%
\subsection{The basic of model building}
In high energy physics, we ask a very simple question: What are the
fundamental laws of Nature? We know that QFT is an adequate tool to
describe Nature, at least at energies we have probed so far. Thus the
question can be stated in a very compact form as: what is the
Lagrangian of nature? The most compact form of the question is
\beq
{\cal L} = ?
\eeq
In order to answer this question we need to provide some axioms or ``rules.''
Our rules are that we ``build'' the Lagrangian by providing the
following three ingredients:
\begin{enumerate}
\item The gauge symmetry of the Lagrangian;
\item The representations of fermions and scalars under the symmetry;
\item The pattern of spontaneous symmetry breaking.
\end{enumerate}
Once these ingredients are provided, we write the most general
renormalizable Lagrangian that is invariant under these symmetries and
provide the required spontaneous symmetry breaking (SSB).

Few remarks are in order about these starting points.
\begin{enumerate}
\item
We also impose Poincare invariance. In a way, this can be identified
as the gauge symmetry of gravity, and thus can be though of part of
the first postulate.
\item
As we already mentioned, we assume QFT. In particular, quantum
mechanics is also an axiom.
\item
We do not impose global symmetries. They are accidental, that is,
they are there only because we do not allow for non renormalizable terms. 
\item
The basic fermion fields are two component Weyl spinors. The basic
scalar fields are complex. The vector fields are introduced to the 
model in order to preserve the gauge symmetry.
\item
Any given model has a finite number of parameters. These parameters
need to be measured before the model can be tested. That is, when we
provide a model, we cannot yet make predictions. Only after an initial
set of measurements are done can we make predictions.
\end{enumerate}

As an example we consider the SM. It is a nice example, mainly because
it describes Nature, and also because the tools we use to construct the
SM are also those we use when constructing its possible extensions.
The SM is defined as follows:
\begin{enumerate}
\item
The gauge symmetry is 
\beq\label{smsym}
G_{\rm SM}=SU(3)_{\rm C}\times SU(2)_{\rm L}\times U(1)_{\rm Y}.
\eeq
\item
There are three fermion generations, each consisting of five 
representations of $G_{\rm SM}$:
\beq\label{ferrep}
Q^I_{Li}(3,2)_{+1/6},\quad U^I_{Ri}(3,1)_{+2/3},\quad
D^I_{Ri}(3,1)_{-1/3},\quad L^I_{Li}(1,2)_{-1/2},\quad E^I_{Ri}(1,1)_{-1}.
\eeq
Our notations mean that, for example, left-handed quarks, $Q_L^I$, are
triplets of $SU(3)_{\rm C}$, doublets of $SU(2)_{\rm L}$ and carry
hypercharge $Y=+1/6$. The super-index $I$ denotes gauge interaction
eigenstates. The sub-index $i=1,2,3$ is the flavor (or generation)
index.  There is a single scalar representation,
\beq\label{scarep}
\phi(1,2)_{+1/2}.
\eeq
\item
The scalar $\phi$ assumes a VEV,
\beq\label{phivev}
\langle\phi\rangle=\pcmatrix{0\cr {v/\sqrt2}\cr},
%\langle\phi\rangle=\begin{pmatrix}0\cr {v\over\sqrt2}\cr \end{pmatrix},
\eeq
which implies that the gauge group is spontaneously broken,
\beq\label{smssb}
G_{\rm SM}\to SU(3)_{\rm C}\times U(1)_{\rm EM}.
\eeq
This SSB pattern is equivalent to requiring that one parameter in
the scalar potential is negative, that is $\mu^2 < 0$, see Eq. (\ref{HiPo}).
\end{enumerate}

The standard model Lagrangian, ${\cal L}_{\rm SM}$, is the most
general renormalizable Lagrangian that is consistent with the gauge
symmetry (\ref{smsym}) and the particle content
(\ref{ferrep}) and (\ref{scarep}). It can be divided to three parts:
\beq\label{LagSM}
{\cal L}_{\rm SM}={\cal L}_{\rm kinetic}+{\cal L}_{\rm Higgs}
+{\cal L}_{\rm Yukawa}.
\eeq
We will learn how to count parameters later, but for now we just
mention that ${\cal L}_{\rm SM}$ has 18 free parameters\footnote{In
fact there is one extra parameter that is related to the vacuum
structure of the strong interaction, $\Theta_{\rm QCD}$. Discussing
this parameter is far beyond the scope of these lectures, and we only
mention it in this footnote in order not to make incorrect
statements.} that we need to determine experimentally. Now we talk a
little about each part of ${\cal L}_{\rm SM}$.

For the kinetic terms, in order to maintain gauge invariance,
one has to replace the derivative with a covariant derivative:
\beq\label{SMDmu}
D^\mu=\partial^\mu+ig_s G^\mu_a L_a+ig W^\mu_b T_b+ig^\prime B^\mu Y.
\eeq
Here $G^\mu_a$ are the eight gluon fields, $W^\mu_b$ the three
weak interaction bosons and $B^\mu$ the single hypercharge boson.
The $L_a$'s are $SU(3)_{\rm C}$ generators (the $3\times3$
Gell-Mann matrices ${1\over2}\lambda_a$ for triplets, $0$ for singlets),
the $T_b$'s are $SU(2)_{\rm L}$ generators (the $2\times2$
Pauli matrices ${1\over2}\tau_b$ for doublets, $0$ for singlets),
and the $Y$'s are the $U(1)_{\rm Y}$ charges. For example, for the
left-handed quarks $Q_L^I$, we have
\beq\label{DmuQL}
{\cal L}_{\rm kinetic}(Q_L)= i{\overline{Q_{Li}^I}}\gamma_\mu
\left(\partial^\mu+{i\over2}g_s G^\mu_a\lambda_a
+{i\over2}g W^\mu_b\tau_b+{i\over6}g^\prime B^\mu\right)Q_{Li}^I,
\eeq
while for the left-handed leptons $L_L^I$, we have
\beq\label{DmuLL}
{\cal L}_{\rm kinetic}(L_L)=i{\overline{L_{Li}^I}}\gamma_\mu 
\left(\partial^\mu+{i\over2}g W^\mu_b\tau_b
-{i\over 2}g^\prime B^\mu\right)L_{Li}^I.
\eeq
This part of the Lagrangian has three parameters, $g$, $g'$ and $g_s$.

The Higgs\footnote{The Higgs mechanism was not first proposed
by Higgs. The first paper suggesting it was by Englert and Brout. It
was independently suggested by Higgs and by Guralnik, Hagen, and
Kibble.}  potential, which describes the scalar self interactions, is
given by:
\beq\label{HiPo}
{\cal L}_{\rm Higgs}=\mu^2\phi^\dagger\phi-\lambda(\phi^\dagger\phi)^2.
\eeq
This part of the Lagrangian involves two parameters, $\lambda$ and
$\mu$, or equivalently, the Higgs mass and its VEV.  The requirement of
vacuum stability tells us that $\lambda>0$. The pattern of spontaneous
symmetry breaking, (\ref{phivev}), requires $\mu^2<0$.

We split the Yukawa part into two, the leptonic and baryonic parts. At
the renormalizable level the lepton Yukawa interactions are given by
\beq\label{Hlint}
-{\cal L}_{\rm Yukawa}^{\rm leptons}=
Y^e_{ij}{\overline {L^I_{Li}}}\phi E^I_{Rj}+{\rm h.c.}.
\eeq
After the Higgs acquires a VEV, these terms lead to charged lepton
masses. Note that the SM predicts massless neutrinos. The Lepton
Yukawa terms involve three physical parameters, which are usually
chosen to be the three charged lepton masses.

The quark Yukawa interactions are given by
\beq\label{Hqint}
-{\cal L}_{\rm Yukawa}^{\rm quarks}=Y^d_{ij}{\overline {Q^I_{Li}}}\phi D^I_{Rj}
+Y^u_{ij}{\overline {Q^I_{Li}}}\tilde\phi U^I_{Rj}+{\rm h.c.}.
\eeq
This is the part where quarks masses and flavor arises, and we will
spend the rest of the lectures on it. For now, just in order to finish
the counting, we mention that the Yukawa interactions for the
quarks are described by ten physical parameters. They can be chosen to
be the six quark masses and the four parameters of the CKM matrix. We
will discuss the CKM matrix at length soon.

The SM has an accidental global symmetry
\beq
U(1)_B \times U(1)_e \times U(1)_\mu \times U(1)_\tau.
\eeq
where $U(1)_B$ is baryon number and the other three $U(1)$s are lepton
family lepton numbers. The quarks carry baryon number, while the
leptons and the bosons do not. We usually normalize it such that the
proton has $B=1$ and thus each quark carries a third unit of baryon
number.  As for lepton number, in the SM each family carries its own
lepton number, $L_e$, $L_\mu$ and $L_\tau$. Total lepton number is a
subgroup of this more general symmetry, that is, the sum of all three
family lepton numbers. In these lectures we concentrate on the quark
sector and therefore we do not elaborate much on the global symmetry
of the lepton sector.

%%%%%%%%%%%%%%%%%%%%%
\subsection{Counting parameters}

Before we go on to study the flavor structure of the SM in detail, we
explain how to identify the number of physical parameter in any
model. The Yukawa interactions of Eq.~(\ref{Hqint}) have many
parameters but some are not physical. That is, there is a basis where
they are identically zero. Of course, it is important to identify the
physical parameters in any model in order to probe and check it.

We start with a very simple example.  Consider a hydrogen atom in a
uniform magnetic field.  Before turning on the magnetic field, the
hydrogen atom is invariant under spatial rotations, which are
described by the $SO(3)$ group.  Furthermore, there is an energy
eigenvalue degeneracy of the Hamiltonian: states with different
angular momenta have the same energy.  This degeneracy is a
consequence of the symmetry of the system.

When magnetic field is added to the system, it is conventional to pick
a direction for the magnetic field without a loss of
generality. Usually, we define the positive $z$ direction to be the
direction of the magnetic field. Consider this choice more carefully.
A generic uniform magnetic field would be described by three real
numbers: the three components of the magnetic field.  The
magnetic field breaks the $SO(3)$ symmetry of the hydrogen atom system
down to an $SO(2)$ symmetry of rotations in the plane perpendicular to
the magnetic field.  The one generator of the $SO(2)$ symmetry is the
only valid symmetry generator now; the remaining two $SO(3)$
generators in the orthogonal planes are broken.  These broken symmetry
generators allow us to rotate the system such that the magnetic field
points in the $z$ direction:
\beq\label{h-magnetic}
  O_{xz} O_{yz} (B_x,B_y,B_z)=(0,0,B'_z),
\eeq 
where $O_{xz}$ and $O_{yz}$ are rotations in the $xz$ and $yz$ planes
respectively. The two broken generators were used to rotate away two
unphysical parameters, leaving us with one physical parameter, the
magnitude of the magnetic field. That is, when turning on the magnetic
field, all measurable quantities in the system depend only on one new
parameter, rather than the na\"ive three. 

The results described above are more generally applicable.
Particularly, they are useful in studying the flavor physics of
quantum field theories.  Consider a gauge theory with matter content.
This theory always has kinetic and gauge terms, which have a certain
global symmetry, $G_f$, on their own.  In adding a potential that
respect the imposed gauge symmetries, the global symmetry may be
broken down to a smaller symmetry group. In breaking the global
symmetry, there is an added freedom to rotate away unphysical
parameters, as when a magnetic field is added to the hydrogen atom
system.  

In order to analyze this process, we define a few quantities.  The
added potential has coefficients that can be described by
$N_\mathrm{general}$ parameters in a general basis.  The global
symmetry of the entire model, $H_f$, has fewer generators than $G_f$
and we call the difference in the number of generators
$N_\mathrm{broken}$.  Finally, the quantity that we would ultimately
like to determine is the number of parameters affecting physical
measurements, $N_\mathrm{phys}$.  These numbers are related by
\beq \label{countrule}
N_\mathrm{phys} = N_\mathrm{general} - N_\mathrm{broken}.  
\eeq 
Furthermore, the rule in (\ref{countrule}) applies separately for both
real parameters (masses and mixing angles) and phases.  A general, $n
\times n$ complex matrix can be parametrized by $n^2$ real parameters
and $n^2$ phases.  Imposing restrictions like Hermiticity or unitarity
reduces the number of parameters required to describe the matrix.  A
Hermitian matrix can be described by $n(n+1)/2$ real parameters and
$n(n-1)/2$ phases, while a unitary matrix can be described by
$n(n-1)/2$ real parameters and $n(n+1)/2$ phases.

The rule given by (\ref{countrule}) can be applied to the standard
model. Consider the quark sector of the model. The kinetic term has a
global symmetry
\beq
G_f=U(3)_Q \times U(3)_U \times U(3)_D.
\eeq
A $U(3)$ has 9 generators (3 real and 6 imaginary), so the total
number of generators of $G_f$ is 27.  The Yukawa interactions defined
in (\ref{Hqint}), $Y^F$ ($F=u,d$), are $3 \times 3$ complex matrices,
which contain a total of 36 parameters (18 real parameters and 18
phases) in a general basis.  These parameters also break $G_f$ down to
the baryon number
\beq
U(3)_Q \times U(3)_U \times U(3)_D 
\to U(1)_B.
\eeq
While $U(3)^3$ has 27 generators, $U(1)_B$ has only one and thus
$N_\mathrm{broken}=26$. This broken symmetry allows us to rotate away
a large number of the parameters by moving to a more convenient basis.
Using (\ref{countrule}), the number of physical parameters should be
given by
\beq 
N_\mathrm{phys} = 36 - 26 = 10.
\eeq
These parameters can be split into real parameters and phases.  The
three unitary matrices generating the symmetry of the kinetic and
gauge terms have a total of 9 real parameters and 18 phases. The
symmetry is broken down to a symmetry with only one phase generator.
Thus,
\beq
N^{(r)}_\mathrm{phys} = 18 - 9 = 9,\qquad N^{(i)}_\mathrm{phys} = 18 - 17
= 1.
\eeq
We interpret this result by saying that of the 9 real parameters, 6
are the fermion masses and three are the CKM matrix mixing angles.
The one phase is the CP-violating phase of the CKM mixing matrix.

In your homework you will count the number of parameters for different models.

%%%%%%%%%%%%%%%%%%%%%%%%%%%%
\subsection{The discrete symmetries of the SM}

Since we are talking a lot about symmetries it is important to recall
the situation with the discrete symmetries, C, P and T. Any local
Lorentz invariant QFT conserves CPT, and in particular, this is also
the case in the SM. CPT conservation also implies that T violation is
equivalent CP violation.

You may wonder why we discuss these symmetries as we are dealing with
flavor. It turns out that in Nature, C, P, and CP violation are
closely related to flavor physics. There is no reason for this to be
the case, but since it is, we study it simultaneously.

In the SM, C and P are ``maximally violated.'' By that we refer to the
fact that both C and P change the chirality of fermion fields. In the
SM the left handed and right handed fields have different gauge
representations, and thus, independent of the values of the parameters
of the model, C and P must be violated in the SM.

The situation with CP is different. The SM can violate CP but it
depends on the values of its parameters. It turns out that the
parameters of the SM that describe Nature violate CP.  The
requirement for CP violation is that there is a physical phase in the
Lagrangian. In the SM the only place where a complex phase can be
physical is in the quark Yukawa interactions. More precisely, in the
SM, CP is violated if and only if
\beq\label{JarCon}
\im{\det[Y^d Y^{d\dagger},Y^u Y^{u\dagger}]}\neq 0.
\eeq

An intuitive explanation of why CP violation is related to complex
Yukawa couplings goes as follows. The Hermiticity of the Lagrangian
implies that ${\cal L}_{\rm Yukawa}$ has pairs of terms in the form
\beq\label{Yukpairs}
Y_{ij}\overline{\psi_{Li}}\phi\psi_{Rj}
+Y_{ij}^*\overline{\psi_{Rj}}\phi^\dagger\psi_{Li}.
\eeq
A CP transformation exchanges the above two operators 
\beq\label{CPoper}
\overline{\psi_{Li}}\phi\psi_{Rj}\ \Leftrightarrow \
\overline{\psi_{Rj}}\phi^\dagger\psi_{Li},
\eeq 
but leaves their coefficients, $Y_{ij}$ and $Y_{ij}^*$, unchanged. This means 
that CP is a symmetry of ${\cal L}_{\rm Yukawa}$ if $Y_{ij}=Y_{ij}^*$.

In the SM the only source of CP violation are Yukawa interactions. It
is easy to see that the kinetic terms are CP conserving. For the
SM scalar sector, where there is a single doublet, this
part of the Lagrangian is also CP conserving.  For extended scalar
sectors, such as that of a two Higgs doublet model, ${\cal L}_{\rm
Higgs}$ can be CP violating.

%%%%%%%%%%%%%%%%%%%%%%%%%%%
\subsection{The CKM matrix} 

We are now equipped with the necessary tools to study the Yukawa
interactions. The basic tool we need is that of basis rotations. There
are two important bases. One where the masses are diagonal, called the
mass basis, and the other where the $W^\pm$ interactions are diagonal,
called the interaction basis. The fact that these two bases are not
the same results in flavor changing interactions. The CKM matrix is
the matrix that rotates between these two bases.

Since most measurements are done in the mass basis, we write the
interactions in that basis.  Upon the replacement
$\re{\phi^0}\to(v+H^0)/\sqrt2$ [see Eq.~(\ref{phivev})], we decompose
the $SU(2)_{\rm L}$ quark doublets into their components:
\beq\label{doublets}
Q^I_{Li}=\pcmatrix{U^I_{Li}\cr D^I_{Li}\cr}.
\eeq
and then the Yukawa interactions, Eq.~(\ref{Hqint}), give rise to mass
terms: 
\beq\label{fermasq}
-{\cal L}_M^q=(M_d)_{ij}{\overline {D^I_{Li}}} D^I_{Rj}
+(M_u)_{ij}{\overline {U^I_{Li}}} U^I_{Rj}+{\rm h.c.}, \qquad
M_q={v\over\sqrt2}Y^q,
\eeq
The mass basis corresponds, by definition, to diagonal mass matrices. We can 
always find unitary matrices $V_{qL}$ and $V_{qR}$ such that
\beq\label{diagMq}
V_{qL}M_q V_{qR}^\dagger=M_q^{\rm diag}\qquad q=u,d,
\eeq
with $M_q^{\rm diag}$ diagonal and real. The quark mass eigenstates
are then identified as
\beq\label{masses}
q_{Li}=(V_{qL})_{ij}q_{Lj}^I,\qquad q_{Ri}=(V_{qR})_{ij}q_{Rj}^I\qquad
q=u,d.
\eeq
The charged current interactions for quarks are the interactions of
the $W^\pm_\mu$, which in the interaction basis are described by
(\ref{DmuQL}). They have a more complicated form in the mass basis:
\beq\label{Wmasq}
-{\cal L}_{W^\pm}^q={g\over\sqrt2}{\overline {u_{Li}}}\gamma^\mu
(V_{uL}V_{dL}^\dagger)_{ij}d_{Lj} W_\mu^++{\rm h.c.}.
\eeq
The unitary $3\times3$ matrix,
\beq\label{VCKM}
V=V_{uL}V_{dL}^\dagger,\qquad
(VV^\dagger={\bf 1}),
\eeq 
is the Cabibbo-Kobayashi-Maskawa (CKM) mixing matrix for quarks. 
%It is
%a unitary $3\times3$ matrix, and thus in general it described by nine
%parameters: three real angles and six phases.
%
As a result of the fact that $V$ is not diagonal, the $W^\pm$ gauge
bosons couple to mass eigenstates quarks of different
generations. Within the SM, this is the only source of flavor changing
quark interactions.

The form of the CKM matrix is not unique. We already counted and
concluded that only one of the phases is physical. This implies that
we can find bases where $V$ has a single phase. This physical phase is
the Kobayashi-Maskawa phase that is usually denoted by $\delta_{\rm
KM}$.
%As we already mentioned, this phase is the single source of CP
%violation in the quark sector of the Standard Model.

There is more freedom in defining $V$ in that we can permute between
the various generations. This freedom is fixed by ordering the up
quarks and the down quarks by their masses, {\it i.e.} 
$(u_1,u_2,u_3)\to(u,c,t)$ and $(d_1,d_2,d_3)\to(d,s,b)$. The elements
of $V$ are therefore written as follows:
\beq\label{defVij}
V=\pcmatrix{V_{ud}&V_{us}&V_{ub}\cr
V_{cd}&V_{cs}&V_{cb}\cr V_{td}&V_{ts}&V_{tb}\cr}.
\eeq

%%%%%%%%%%%%%%%%%%%%%%%%%%
%\subsection{The unitarity triangles}

The fact that there are only three real and one imaginary physical
parameters in $V$ can be made manifest by choosing an explicit
parametrization.  For example, the standard parametrization, used by
the Particle Data Group (PDG)~\cite{pdg}, is given by
\beq\label{stapar}
V=\pcmatrix{c_{12}c_{13}&s_{12}c_{13}&
s_{13}e^{-i\delta}\cr 
-s_{12}c_{23}-c_{12}s_{23}s_{13}e^{i\delta}&
c_{12}c_{23}-s_{12}s_{23}s_{13}e^{i\delta}&s_{23}c_{13}\cr
s_{12}s_{23}-c_{12}c_{23}s_{13}e^{i\delta}&
-c_{12}s_{23}-s_{12}c_{23}s_{13}e^{i\delta}&c_{23}c_{13}\cr},
\eeq
where $c_{ij}\equiv\cos\theta_{ij}$ and
$s_{ij}\equiv\sin\theta_{ij}$. The three $\sin\theta_{ij}$ are the
three real mixing parameters while $\delta$ is the Kobayashi-Maskawa
phase. Another parametrization is the Wolfenstein
parametrization where the four mixing parameters are
$(\lambda,A,\rho,\eta)$ where $\eta$ represents the CP violating
phase. The Wolfenstein parametrization is an expansion in
the small parameter, $\lambda=|V_{us}|\approx 0.22$. To $O(\lambda^3)$
the parametrization is given by
%\beq\label{wolpar}
%V=\pcmatrix{
%1-\frac12\lambda^2-\frac18\lambda^4 & \lambda & A\lambda^3(\rho-i\eta)\cr
%-\lambda +\frac12A^2\lambda^5[1-2(\rho+i\eta)] &
%1-\frac12\lambda^2-\frac18\lambda^4(1+4A^2) & A\lambda^2 \cr
%A\lambda^3[1-(1-\frac12\lambda^2)(\rho+i\eta)]&-A\lambda^2+\frac12A\lambda^4[1-2(\rho+i\eta)]
%& 1-\frac12A^2\lambda^4 \cr}\; .
%\eeq
\beq\label{wolpar}
V=\pcmatrix{
1-\frac12\lambda^2 & \lambda & A\lambda^3(\rho-i\eta)\cr
-\lambda  & 1-\frac12\lambda^2 & A\lambda^2 \cr
A\lambda^3(1-\rho-i\eta)&-A\lambda^2 & 1 \cr}\; .
\eeq
We will talk in detail about how to measure the CKM parameters. For
now let us mention that the Wolfenstein parametrization is a good
approximation to the actual numerical values. That is, the CKM matrix
is very close to a unit matrix with off diagonal terms that are
small. The order of magnitude of each element can be read from the
power of $\lambda$ in the Wolfenstein parametrization.

Various parametrizations differ in the way that the freedom of phase
rotation is used to leave a single phase in $V$. One can define,
however, a CP violating quantity in $V$ that is independent of the
parametrization. This quantity, the Jarlskog invariant, $J_{\rm CKM}$,
is defined through
\beq\label{defJ}
\im{V_{ij}V_{kl}V_{il}^*V_{kj}^*}=
J_{\rm CKM}\sum_{m,n=1}^3\epsilon_{ikm}\epsilon_{jln},\qquad (i,j,k,l=1,2,3).
\eeq
In terms of the explicit parametrizations given above, we have
\beq\label{parJ}
J_{\rm CKM}=c_{12}c_{23}c_{13}^2s_{12}s_{23}s_{13}\sin\delta\approx
\lambda^6 A^2\eta.
\eeq

The condition (\ref{JarCon}) can be translated to the language of the
flavor parameters in the mass basis.  Then we see that a necessary and
sufficient condition for CP violation in the quark sector of the SM
(we define $\Delta m^2_{ij}\equiv m_i^2-m_j^2$):
\beq\label{jarconmas}
\Delta m^2_{tc}\Delta m^2_{tu}\Delta m^2_{cu}\Delta m^2_{bs}\Delta m^2_{bd}
\Delta m^2_{sd}J_{\rm CKM}\neq0.
\eeq
Equation (\ref{jarconmas}) puts the following requirements on the SM in order
that it violates CP:
\begin{enumerate}
\item
Within each quark sector, there should be no mass degeneracy;
\item
None of the three mixing angles should be zero or $\pi/2$;
\item
The phase should be neither 0 nor $\pi$.
\end{enumerate}

A very useful concept is that of the unitarity triangles. The
unitarity of the CKM matrix leads to various relations among the
matrix elements, for example, 
\beq
\sum_i V_{id}V_{is}^*=0.
\eeq
There are six such relations and they require the sum of three complex
quantities to vanish. Therefore, they can be geometrically represented
in the complex plane as a triangle and are called ``unitarity
triangles".  It is a feature of the CKM matrix that all unitarity
triangles have equal areas. Moreover, the area of each unitarity
triangle equals $|J_{\rm CKM}|/2$ while the sign of $J_{\rm CKM}$
gives the direction of the complex vectors around the triangles.

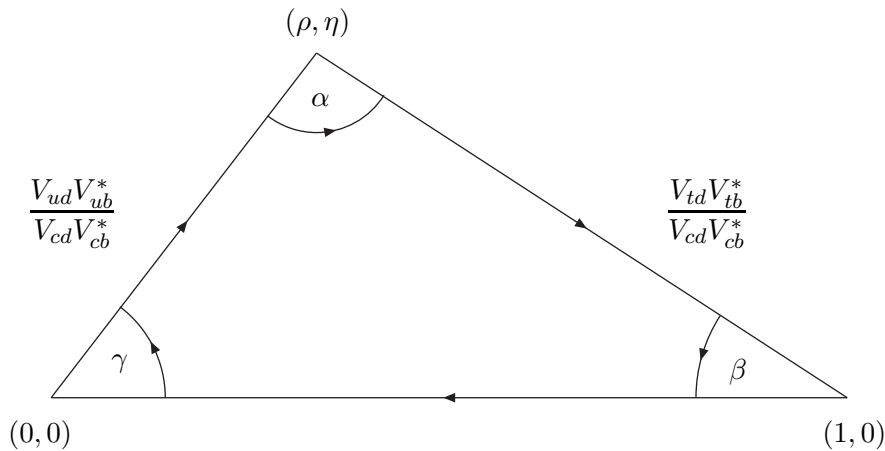
\begin{figure}[t]
\begin{center}
\begin{picture}(350,160)(0,20)
\ArrowLine(20,30)(120,160)
\ArrowLine(120,160)(320,30)
\ArrowLine(320,30)(20,30)
\Text(28,100)[]{$\displaystyle{\frac{V_{ud} V_{ub}^*}{V_{cd} V_{cb}^*}}$}
\Text(268,100)[]{$\displaystyle{\frac{V_{td} V_{tb}^*}{V_{cd} V_{cb}^*}}$}
%\Text(150,10)[]{$V_{cd} V_{cb}^*$}
\Text(121,172)[]{$(\rho,\eta)$}
\Text(122,143)[]{$\alpha$} 
\ArrowArc(120,160)(30,232,328)
\Text(280,40)[]{$\beta$}
\ArrowArc(320,30)(57,147,180)
\Text(46,43)[]{$\gamma$}
\ArrowArc(20,30)(43,0,53)
\Text(16,16)[]{$(0,0)$}
\Text(324,16)[]{$(1,0)$}
\end{picture}
\end{center}
\caption[y]{The unitarity triangle. \label{fg:tri}}
\end{figure}

One of these triangles has sides roughly the same length, is
relatively easy to probe, and is corresponds to the relation
\beq\label{Unitdb}
V_{ud}V_{ub}^*+V_{cd}V_{cb}^*+V_{td}V_{tb}^*=0.
\eeq
For these reasons, the term ``the unitarity triangle'' is reserved for
Eq.~(\ref{Unitdb}).  We further define the rescaled unitarity
triangle. It is derived from (\ref{Unitdb}) by choosing a phase
convention such that $(V_{cd}V_{cb}^*)$ is real and dividing the
lengths of all sides by $|V_{cd}V_{cb}^*|$. The rescaled unitarity
triangle is similar to the unitarity triangle. Two vertices of the
rescaled unitarity triangle are fixed at (0,0) and (1,0). The
coordinates of the remaining vertex correspond to the Wolfenstein
parameters $(\rho,\eta)$. The unitarity triangle is shown in
Fig. \ref{fg:tri}.

The lengths of the two complex sides are
\beq\label{RbRt}
R_u\equiv\left|{V_{ud}V_{ub}\over V_{cd}V_{cb}}\right|
=\sqrt{\rho^2+\eta^2},\qquad
R_t\equiv\left|{V_{td}V_{tb}\over V_{cd}V_{cb}}\right|
=\sqrt{(1-\rho)^2+\eta^2}.
\eeq
The three angles of the unitarity triangle are defined as follows:
\beq\label{abcangles}
\alpha\equiv\arg\left[-{V_{td}V_{tb}^*\over V_{ud}V_{ub}^*}\right],\qquad
\beta\equiv\arg\left[-{V_{cd}V_{cb}^*\over V_{td}V_{tb}^*}\right],\qquad
\gamma\equiv\arg\left[-{V_{ud}V_{ub}^*\over V_{cd}V_{cb}^*}\right].
\eeq
They are physical quantities and can be independently measured, as we will
discuss later. Another commonly used notation is $\phi_1=\beta$,
$\phi_2=\alpha$, and $\phi_3=\gamma$. Note that in the standard
parametrization $\gamma=\delta_{KM}$.

%%%%%%%%%%%%%%%%%%%%%%%%%%%%%%%%%%%%
\subsection{FCNCs}

So far we have talked about flavor changing charged currents that are
mediated by the $W^\pm$ bosons. In the SM, this is the only source of
flavor changing interaction and, in particular, of generation changing
interaction. There is no fundamental reason why there cannot be
Flavor Changing Neutral Currents (FCNCs). After all, two interactions
of flavor changing charged current result in a neutral current
interaction. Yet, experimentally we see that FCNCs processes are
highly suppressed.

This is a good place to pause and open your PDG.\footnote{It goes
without saying that every student in high energy physics must have
the PDG~\cite{pdg}. If, for some reason you do not have one, order it
now. It is free and has a lot of important stuff. Until you get it,
you can use the online version at pdg.lbl.gov.}  Look, for example, at
the rate for the neutral current decay, $K_L \to \mu^+\mu^-$, and compare
it to that of the charged current decay, $K^+ \to \mu^+ \nu$. You see
that the $K_L$ decay rate is much smaller. It is a good idea at this
stage to browse the PDG a bit more and see that the same pattern is
found in $D$ and $B$ decays.

The fact that the data show that FCNCs are highly suppressed implies
that any model that aims to describe Nature must have a mechanism to
suppress FCNCs. The SM's way to deal with the data is to make sure there
are no tree level FCNCs. In the SM, FCNCs are mediated only at the
loop level and are therefore suppressed (we discuss the exact amount
of suppression below). Next we explain why in the SM all neutral
current interactions are flavor conserving at the tree level.

Before that, we make a short remark. We often talk about non-diagonal
couplings, diagonal couplings and universal couplings. Universal
couplings are diagonal couplings with the same strength. An important
point to recall is that universal couplings are diagonal in any
basis. Non-universal diagonal couplings, in general, become
non-diagonal after a basis rotation.

There are four types of neutral bosons in the SM that could mediate
tree level neutral currents. They are the gluons, photon, Higgs and
$Z$ bosons. We study each of them in turn, explain what is required in order
to make their couplings diagonal in the mass basis, and how this
requirement is fulfilled in the SM.

We start with the massless gauge bosons: the gluons and photon. For
them, tree level couplings are always diagonal, independent of the
details of the theory. The reason is that these gauge bosons
correspond to exact gauge symmetries. Thus, their couplings to the
fermions arise from the kinetic terms. When the kinetic terms are
canonical, the couplings of the gauge bosons are universal and, in
particular, flavor conserving. In other words, gauge symmetry plays a
dual role: it guarantees that the gauge bosons are massless and that
their couplings are flavor universal.

Next we move to the Higgs interactions. The reason that the Higgs
couplings are diagonal in the SM is because its couplings to
fermions are aligned with the mass matrix. The reason is that both the
Higgs coupling and the mass matrix are proportional to the same Yukawa
couplings. To see that this is the case we consider the Yukawa
interactions (\ref{Hqint}). After inserting
$\re{\phi^0}\to(v+h)/\sqrt2$ and keeping both the fermion masses and
Higgs fermion interaction terms we get
\beqa
-{\cal L}_{\rm Yukawa}^{\rm quarks}&=&
Y^d_{ij}{\overline {Q^I_{Li}}}\phi D^I_{Rj}
+Y^u_{ij}{\overline {Q^I_{Li}}}\tilde\phi U^I_{Rj} \nonumber \\
&=&
{Y^d_{ij}\over \sqrt{2}}\left({\overline {D^I_{Li}}} D^I_{Rj}\right)(v+h)
+{Y^u_{ij}\over \sqrt{2}}\left({\overline {U^I_{Li}}}
U^I_{Rj}\right)(v+h).
\eeqa
Diagonalizing the mass matrix, we get the interaction in the physical basis
\beq
M_d\left({\overline {D_{Li}}} D_{Rj}\right)(v+h)
+M_u\left({\overline {U_{Li}}} U_{Rj}\right)(v+h).
\eeq
Clearly, since everything is proportional to $(v+h)$, the interaction
is diagonalized together with the mass matrix.

This special feature of the Higgs interaction is tightly related to
the facts that the SM has only one Higgs field and that the only
source for fermion masses is the Higgs VEV. In models where there are
additional sources for the masses, that is, bare mass terms or more
Higgs fields, diagonalization of the mass matrix does not
simultaneously diagonalize the Higgs interactions.  In general, there
are Higgs mediated FCNCs in such models. In your homework you will
work out an example of such models.

Last, we discuss $Z$-mediated FCNCs. The coupling for the $Z$ to
fermions is proportional to $T_3-q \sin^2\theta_W$ and in the
interaction basis the $Z$ couplings to quarks are given by
\beqa
-{\cal L}_Z&=&{g \over \cos\theta_W} \left[
{\overline {u^I_{Li}}}\gamma^\mu\left({1\over 2}-{2\over 3}
\sin^2\theta_W\right) u^I_{Li} +
{\overline {u^I_{Ri}}}\gamma^\mu\left(-{2\over 3}
\sin^2\theta_W\right) u^I_{Ri} +\right. \nonumber \\ &&
\left. {\overline {d^I_{Li}}}\gamma^\mu\left(-{1\over 2}+{1\over 3}
\sin^2\theta_W\right) d^I_{Li} +
{\overline {d^I_{Ri}}}\gamma^\mu\left({1\over 3}
\sin^2\theta_W\right) d^I_{Ri}\right]Z_\mu +{\rm h.c.}.
\eeqa
In order to demonstrate the fact that there are no FCNCs let us
concentrate only on the left handed up-type quarks.  Moving to the
mass eigenstates we find 
\beqa
-{\cal L}_Z&=&{g \over \cos\theta_W} 
\left[ {\overline {u_{Li}}}\;\left(V_{uL}\right)_{ik}
\;\gamma^\mu\left({1\over 2}-{2\over 3} \sin^2\theta_W\right)
\left(V_{uL}^\dagger\right)_{kj} u_{Lj}\right]\,,\nonumber \\
&=&
{g \over \cos\theta_W} 
\left[ {\overline {u_{Li}}}\,
\gamma^\mu\left({1\over 2}-{2\over 3} \sin^2\theta_W\right)
u_{Li}\right]
\eeqa
where in the last step we used
\beq
V_{uL}V_{uL}^\dagger=1.
\eeq
We see that the interaction is universal and diagonal in flavor. It is
easy to verify that this holds for the other types of quarks.  
Note the difference between the neutral and the charged currents cases.
In the neutral current case we insert $V_{uL}V_{uL}^\dagger=1$.  This is in
contrast to the charged current interactions where the insertion is
$V_{uL}V_{dL}^\dagger$, which in general is not equal to the identity
matrix.

The fact that there are no FCNCs in $Z$-exchange is due to some
specific properties of the SM. That is, we could have $Z$-mediated
FCNCs in simple modifications of the SM. The general condition for the
absence of tree level FCNCs is as follows. In general, fields can mix
if they belong to the same representation under all the {\em unbroken}
generators. That is, they must have the same spin, electric charge and
SU(3)$_C$ representation.  If these fields also belong to the same
representation under the {\em broken} generators their couplings to
the massive gauge boson is universal. If, however, they belong to 
different representations under the broken generators, their couplings
in the interaction basis are diagonal but non-universal. These
couplings become non-diagonal after rotation to the mass basis.

In the SM, the requirement mention above for the absence of
$Z$-exchange FCNCs is satisfied. That is, all the fields that belong
to the same representation under the unbroken generators also belong
to the same representation under the broken generators.  For example,
all left handed quarks with electric charge $2/3$ also have the same
hypercharge ($1/6$) and they are all an up component of a double of
$SU(2)_L$ and thus have $T_3=1/2$. This does not have to be the
case. After all, $Q=T_3+Y$, so there are many ways to get quarks with
the same electric charge.  In your homework, you will work out the
details of a model with non-standard representations and see how it
exhibits $Z$-exchange FCNCs.

%%%%%%%%%%%%%%%%%%%%%%%%%%%%
\subsection{Homework}

\que{Global symmetries}

We talked about the fact that global symmetries are accidental in the
SM, that is, that they are broken once non-renormalizable terms are
included. Write the lowest dimension terms that break each of the
global symmetries of the SM.

\vspace*{3mm}
\que{Extra generations counting}

Count the number of physical flavor parameters in an extended SM with
$n$ generations. Show that such a model has $n(n+3)/2$ real parameters
and $(n-1)(n-2)/2$ complex parameters. Identify the real parameters as
masses and mixing angles and determine how many mixing angles there
are.

\vspace*{3mm}
\que{Exotic light quarks}

We consider a model with the gauge symmetry $SU(3)_C \times SU(2)_L
\times U(1)_Y$ spontaneously broken by a single Higgs doublet into
$SU(3)_C \times U(1)_{EM}$.  The quark sector, however, differs from
the standard model one as it consists of three quark flavors, that
is, we do not have the $c$, $b$ and $t$ quarks.  The quark
representations are non-standard.  Of the left handed quarks,
$Q_L=(u_L,d_L)$ form a doublet of $SU(2)_L$ while $s_L$ is a singlet.
All the right handed quarks are singlets. All color representations
and electric charges are the same as in the standard model.

\ben
\item
Write down (a) the gauge interactions of the quarks with the charged
$W$ bosons (before SSB); (b) the Yukawa interactions (before SSB); (c)
the bare mass terms (before SSB); (d) the mass terms after SSB.

\item
Show that there are five physical flavor parameters in this model. How
many are real and how many imaginary?
Is there CP violation in this model?
Separate the five into masses, mixing angles and phases. 

\item
Write down the gauge interactions of the quarks with the $Z$ boson in
both the interaction basis and the mass basis. (You do not have to
rewrite terms that do not change when you rotate to the mass
basis. Write only the terms that are modified by the rotation to the
mass basis.)  Are there, in general, tree level $Z$ exchange FCNCs?
(You can assume CP conservation from now on.)

\item
Are there photon and gluon mediated FCNCs? Support your answer by an
argument based on symmetries.

\item
Are there Higgs exchange FCNCs?

\item
Repeat the question with a somewhat different model, where the only
modification is that two of the right handed quarks, $Q_R=(u_R,d_R)$,
form a doublet of $SU(2)_L$. Note that there is one relation between
the real parameters that makes the parameter counting a bit tricky.

\een

\que{Two Higgs doublet model}

Consider the two Higgs doublet model (2HDM) extension of the SM. 
In this model, we add a Higgs doublet to the SM fields.
Namely, instead of the one Higgs field of the SM we now have two, denoted by
$\phi_1$ and $\phi_2$. For simplicity you can work with two generations
when the third generation is not explicitly needed.
\ben
\item
Write down (in a matrix notation) the most general Yukawa potential
of the quarks.

\item
Carry out the diagonalization procedure for such a model. Show that 
the $Z$ couplings are still flavor diagonal. 

\item
In general, however, there are FCNCs in this model mediated by 
the Higgs bosons. To show that, write the Higgs fields as 
$\Re(\phi_i) = v_i + h_i$
where $i=1,2$ and $v_i \ne 0$ is the VEV of $\phi_i$, and define 
$\tan\beta=v_2/v_1$.
Then, write down the Higgs--fermion interaction terms in the mass basis.
Assuming that there is no mixing between the Higgs fields, you should
find a non-diagonal Higgs fermion interaction terms. 

\een

%%%%%%%%%%
%%%%%%%%%%
%%%%%%%%%%
\section{Probing the CKM}

Now that we have an idea about flavor in general and in the SM in
particular, we are ready to compare the standard model predictions
with data. While we use the SM as an example, the tools and
ideas are applicable to a large number of theories.

%\subsection{Determining the CKM}

The basic idea is as follows. In order to check a model we first have
to determine its parameters and then we can probe it. When considering
the flavor sector of the SM, this implies that we first have to
measure the parameters of the CKM matrix and then check the
model. That is, we can think about the first four measurements as
determining the CKM parameters and from the fifth measurements on we are
checking the SM. In practice, however, we look for many independent
ways to determine the parameters. The SM is checked by looking for
consistency among these measurements. Any inconsistency is a signal of
new physics.\footnote{The term ``new physics'' refers to any model
that extends the SM. Basically, we are eager to find indications for
new physics and determine what that new physics is. At the end of the
lectures we expand on this point.}

There is one major issue that we need to think about: how precisely
can the predictions of the theory be tested? Our ability to test
any theory is bounded by these precisions. There are two kinds of
uncertainties: experimental and theoretical. There are many sources of
both kinds, and a lot of research has gone into trying to overcome them in
order to be able to better probe the SM and its extensions.

We do not elaborate on experimental details. We just make one general
point. Since our goal is to probe the small elements of the CKM, we
have to measure very small branching ratios, typically down to
$O(10^{-6})$. To do that we need a lot of statistics and a superb
understanding of the detectors and the backgrounds.

As for theory errors, there is basically one player here: QCD, or,
using its mighty name, ``the strong interaction.'' Yes, it is strong,
and yes, it is a problem for us. Basically, we can only deal with
weakly coupled forces. The use of perturbation theory is so
fundamental to our way of doing physics. It is very hard to deal with
phenomena that we cannot use perturbation theory to describe.

In practice the problem is that our theory is given in terms of
quarks, but measurements are done with hadrons. It is far from trivial
to overcome this gap. In particular, it becomes hard when we are
looking for high precision.  There are basically two ways to overcome
the problem of QCD. One way is to find observables for which the
needed hadronic input can be measured or eliminated. The other way is
to use approximate symmetries of QCD, in particular, isospin,
$SU(3)_F$ and heavy quark symmetries. Below we only mention how these
are used without getting into much detail.

%%%%%%%%%%%%%%%%%%%%%%%%%%
\subsection{Measuring the CKM parameters}

When we attempt to determine the CKM parameters we talk about two
classifications. One classification is related to what we are trying
to extract:
\ben
\item
Measure magnitudes of CKM elements or, equivalently, sides of the
unitarity triangle;
\item
Measure phases of CKM elements or, equivalently, angles of the
unitarity triangle;
\item
Measure combinations of magnitudes and phases.
\een
The other classification is based on the physics, in particular, we
classify based on the type of amplitudes that are involved:
\ben
\item
Tree level amplitudes.  Such measurements are also referred to as
``direct measurements;''
\item
Loop amplitudes.  Such measurements are also referred to as
``indirect measurements;''
\item
Cases where both tree level and loop amplitude are involved.
\een

There is no fundamental difference between direct and indirect
measurement. We make the distinction since direct measurements are
expected to be almost model independent. Most extensions of the SM
have a special flavor structure that suppresses flavor changing
couplings and have a very small effect on processes that are large in
the SM, which are tree level processes. On the contrary, new physics
can have large effect on processes that are very small in the SM,
mainly loop processes. Thus, we refer to loop amplitude measurements as
indirect measurements.

\subsection{Direct measurements}

In order to determine the magnitudes of CKM elements, a number of
sophisticated theoretical and experimental techniques are needed, the
complete discussion of which is beyond the scope of these
lectures. Instead, we give one example, the determination of
$|V_{cb}|$ and, hopefully, you will find the time to read about direct
determinations of other CKM parameters in one of the reviews such as
the PDG or~\cite{Hocker:2006xb}.

The basic idea in direct determination of CKM elements is to use the
fact that the amplitudes of semi leptonic tree level decays are
proportional to one CKM element. In the case of $b \to c \ell \bar\nu$
it is proportional to $V_{cb}$. (While the diagram is not plotted
here, it is a good time to pause and see that you can plot it and see
how the dependence on the CKM element enters.) Experimentally, it
turns out that $|V_{ub}| \ll |V_{cb}|$. Therefore we can neglect the
$b\to u \ell \bar\nu$ decays and use the full semileptonic $B$ decays
data set to measure $|V_{cb}|$ without the need to know the hadronic
final state.

The way to overcome the problem of QCD is to use heavy quarks symmetry
(HQS). We do not discuss the use of HQS in detail here.  We just
mention that the small expansion parameter is $\Lambda_{QCD}/m_b$. The
CKM element $|V_{cb}|$ can be extracted from inclusive and exclusive
semileptonic $B$ decays.

In the inclusive case, the problem is that the calculation is done
using the $b$ and $c$ quarks. In particular, the biggest uncertainty
is the fact that at the quark level the decay rate scales like
$m_b^5$. The definition of the $b$ quark mass, as well as the
measurements of it, is complicated: How can we define a mass to a
particle that is never free? All we can define and measure very
precisely is the $B$ meson mass.\footnote{There is an easy way to
remember the mass of the $B$ meson that is based on the fact that it
is easier to remember two things than one. I often ask people how many
feet there are in one mile, and they do not know the answer. Most of
them also do not know the mass of the $B$ meson in MeV. It is rather
amusing to note that the answer is, in fact, the same, 5280.} Using an
operator product expansion (OPE) together with the heavy quark
effective theory, we can expand in the small parameter and get a
reasonable estimate of $|V_{cb}|$. The point to emphasize is that this
is a controllable expansion, that is, we know that
\beq
\Gamma (b \to c \ell \bar\nu) = 
\Gamma (B \to X_c \ell \bar\nu) \left(1+ \sum_n a_n\right),
\eeq
such that $a_n$ is suppressed by $(\Lambda_{QCD}/m_B)^n$. In principle
we can calculate all the $a_n$ and get a very precise prediction. It
is helpful that $a_1=0$. The calculation has been done for $n=2$ and
$n=3$.

The exclusive approach overcomes the problem of the $b$ quark mass by
looking at specific hadronic decays, in particular, $B \to D \ell
\bar\nu$ and $B \to D^* \ell \bar\nu$. Here the problem is that the
decay cannot be calculated in terms of quarks: it has to be done in
terms of hadrons. This is where using ``form factors'' is useful as we
now explain. The way to think about the problem is that a $b$ field
creates a free $b$ quark or annihilates a free anti-$b$ quark. Yet,
inside the meson the $b$ is not free. Thus the operator that we care
about, $\bar c \gamma_\mu b$, is not directly related to annihilating
the $b$ quark inside the meson. The mismatch is parametrized by form
factors. The form factors are functions of the momentum transfer. In
general, we need some model to calculate these form factors, as they
are related to the strong interaction. In the $B$ case, we can use
HQS, which tells us that in the limit $m_B \to \infty$ all the form
factors are universal, and are given by the (unknown) Isgur-Wise
function. The fact that we know something about the form factors makes
the theoretical errors rather small, below the $5\%$ level.

Similar ideas are used when probing other CKM elements. For example,
in $\beta$-decay the $d \to u e \bar\nu$ decay amplitude is
proportional to $V_{ud}$. Here the way to overcome QCD is by using
isospin where the expansion parameter is $m_q/\Lambda_{QCD}$ with
$q=u,d$. Another example is $K$-decay, $s \to u e \bar\nu \propto
V_{us}$. In that case, in addition to isospin, flavor SU(3) is used
where we assume that the strange quark is light. In some cases, this is a
good approximation, but not as good as isospin.

Direct measurements have been used to measure the magnitude of seven
out of the nine CKM matrix components. The two exceptions are
$|V_{ts}|$ and $|V_{td}|$. The reason is that the world sample of top
decays is very small, and moreover, it is very hard to determine the flavor
of the light quark in top decay. These two elements are best probed
using loop processes, as we discuss next.

%%%%%%%%%%%%%%%%%%
\subsection{Indirect measurements}

The CKM dependence of decay amplitudes involved in direct measurements
of the CKM elements is simple. The amplitudes are tree level with one
internal $W$ propagator. In the case of semileptonic decays, the
amplitude is directly proportional to one CKM matrix element.

The situation with loop decays is different. Usually we concentrate on
FCNC\footnote{In the first lecture we proved that in the SM there are
no tree-level FCNCs. Why do we talk about FCNCs here? I hope the
answer is clear.} processes at the one loop level. Since the loop
contains an internal $W$ propagator, we gain sensitivity to CKM
elements. The sensitivity is always to a combination of CKM
elements. Moreover, there are several amplitudes with different
internal quarks in the loop. These amplitudes come with different
combinations of CKM elements. The total amplitude is the sum of these
diagrams, and thus it has a non trivial dependence on combination of
CKM elements.

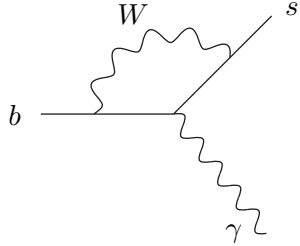
\begin{figure}
\begin{center}
\begin{picture}(210,100)(-40,25)
\Line(0,80)(50,80)
\Line(50,80)(87,117)
%\Line(85,35)(85,-15)
%\Line(85,35)(135,35)
\Photon(50,80)(85,35) 3 5
\PhotonArc(50,80)(30,45,180) 3 5
\Text(-10,80)[]{$b$}
\Text(73,35)[]{$\gamma$}
%\Text(72,5)[]{$\overline q$}
\Text(95,120)[]{$s$}
\Text(35,118)[]{$W$}
\end{picture}
\end{center}
\caption{One of the $b\to s \gamma$ amplitudes\label{bsgamma-fig}}
\end{figure}
 
As an example consider one of the most interesting loop induced decay,
$b \to s \gamma$. There are several amplitudes for this decay. One of
them is plotted in Fig.~\ref{bsgamma-fig}. (Try to plot the others
yourself. Basically the difference is where the photon line goes out.)
Note that we have to sum over all possible internal quarks. Each set
of diagrams with a given internal up-type quarks, $u_i$, is
proportional to $V_{ib}V_{is}^*$. It can further depend on the mass of
the internal quark. Thus, we can write the total amplitude as
\beq \label{a-bsg}
A(b \to s \gamma)\propto \sum_{i=u,c,t} f(m_i) V_{ib}V_{is}^*.
\eeq
While the expression in (\ref{a-bsg}) looks rather abstract, we can
gain a lot of insight into the structure of the amplitude by recalling
that the CKM matrix is unitary. Using
\beq \label{uniqqq}
\sum_{i=u,c,t}  V_{ib}V_{is}^*=0,
\eeq
we learn that the contribution of the $m_i$ independent term in $f$
vanishes. Explicit calculation shows that $f(m_i)$ grows with $m_i$
and, if expanding in $m_i/m_W$, that the leading term scales like $m_i^2$.

The fact that in loop decays the amplitude is proportional to
$m_i^2/m_W^2$ is called the GIM mechanism. Historically, it was the
first theoretical motivation of the charm quark. Before the charm was
discovered, it was a puzzle that the decay $K_L \to \mu^+
\mu^-$ was not observed. The GIM mechanism provided an answer. The
fact that the CKM is unitary implies that this process is a one loop
process and there is an extra suppression of order $m_c^2/m_W^2$ to
the amplitude. Thus, the rate is tiny and very hard to observe.

The GIM mechanism is also important in understanding the finiteness of
loop amplitudes. Any one loop amplitude corresponding to decay where
the tree level amplitude is zero must be finite. Technically, this can
be seen by noticing that if it were divergence, a counter term at
tree-level would be needed, but that cannot be the case if the
tree-level amplitude vanishes. The amplitude for $b\to s
\gamma$ it is naively log divergent. (Make sure you do the counting
and see it for yourself.) Yet, it is only the $m_i$ independent term
that diverges. The GIM mechanism is here to save us as it
guarantees that this term is zero. The $m_i$ dependent term is finite,
as it should be.

One more important point about the GIM mechanism is the fact that the
amplitude is proportional to the mass squared of the internal
quark. This implies that the total amplitude is more sensitive to
couplings of the heavy quarks. In $B$ decays, the heaviest internal
quark is the top quark. This is the reason that $b\to s\gamma$ is
sensitive to $|V_{ts}V_{tb}|$. This is a welcome feature since, as we
mentioned before, these elements are hard to probe directly.

In one loop decays of kaons, there is a ``competition'' between the
internal top and charm quarks. The top is heavier, but the CKM
couplings of the charm are larger. Numerically, the charm is the winner,
but not by a large margin. Check for yourself.

As for charm decay, since the tree level decay amplitudes are large,
and since there is no heavy internal quark, the loop decay amplitudes
are highly suppressed. So far the experimental bounds on various
loop-mediated charm decays are far above the SM predictions. As an
exercise, try to determine which internal quark dominates the one loop
charm decay.

\subsection{Homework}

\que{Direct CKM measurements from $D$ decays}

The ratio of CKM elements
\beq
r\equiv {|V_{cd}| \over |V_{cs}|}
\eeq
can be estimated assuming SU(3) flavor symmetry. The idea is that in
the SU(3) limit the pion and the kaon have the same mass and the same
hadronic matrix elements. 
\begin{enumerate}
\item
Construct a ratio of semileptonic $D$ decays that can be use to
measure the ratio $r$.
\item
We usually expect SU(3) breaking effects to be of the order
$m_s/\Lambda_{QCD} \sim 20\%$. Compare the observable you constructed
to the actual measurement and estimate the SU(3) breaking effect.
\end{enumerate}

\que{The GIM mechanism: $b\to s \gamma$ decay}

\begin{enumerate}
\item
Explain why $b\to s \gamma$ is a loop decay and
draw the one loop diagrams in the SM.
\Answer{The photon couplings is always diagonal so it must be a loop
decay. The diagrams are a simple one loop with $W$ and up-type quark
in the loop. The photon then can go out of any of the lines, that is,
$b$, $s$, $W$ and $u_i$.}
\item
Naively, these diagrams diverge. Show this.
\Answer{Lets look at the diagram where the photon is connected to the
internal up type quark. Consider the diagram with internal top. The
diagram is roughly proportional to
\beq
 V_{ts}V^*{tb} \int d^4 k\, {(\kslash - m_t)^2 \over (k^2-m_t^2)^2(k^2-m_W^2)}
\eeq
The term that go like $\kslash$ vanish (as it is antisymmetric) and
thus in the UV we end up with
\beq
 V_{ts}V^*{tb} \int d^4 k\, {k^2+m_t^2 \over k^6}
\eeq
and it is log divergent.
}
\item
Once we add all the diagrams and make use of the CKM unitarity, we get
a finite result. Show that the UV divergences cancel (that is, put all
masses the same and show that the answer is zero).
\Answer{
When we add it all the UV part is
\beq
\sum_i V_{is}V^*{ib} \int d^4 k\, {k^2+m_i^2 \over k^6}
\eeq
Using unitarity we see that the divergent term vanish (as it does not
depend on the external mass) The only one that is there is the one
that proportional to $m_i^2$, and it is finite.}
\item
We now add a vector-like pair of down type quarks to the SM which we
denote by $b'$
\beq
b'_R(3,1)_{-1/3}, \qquad b'_L(3,1)_{-1/3}.
\eeq
Show that in that model Eq.~(\ref{uniqqq}) is not valid anymore, that is,
\beq
\sum_{i=u,c,t}  V_{ib}V_{is}^*\ne 0,
\eeq
and that we have a $Z$ exchange tree level FCNCs in the down sector.
(The name ``vector-like'' refers to the case where the left and right
handed fields have the same representation under all gauge groups. This
is in contrast to a chiral pair where they have different
representations. All the SM fermions are chiral.)
\item
As we argued, in any model we cannot have $b \to s \gamma$ at tree
level. Thus, in the model with the vector-like quarks, the one
loop diagrams must also be finite. Yet, in the SM we used Eq.~(\ref{uniqqq}) to
argue that the amplitude is finite, but now it is not valid. Show that
the amplitude is finite also in this case. (Hint: When you have an
infinite result that should be finite the reason is usually that
there are more diagrams that you forgot.)
\Answer{
Now there is also a one loop diagram with $Z$ in the loop. The
$bs$ FCNC coupling of the $Z$ is proportional to
$V_{is}V_{ib}^*$. Thus, once you include this diagram we regain
unitarity and the sum is finite.}
\end{enumerate}

%%%%%%%%%%%%%%%%%%%%%%%%%%
\section{Meson mixing}

Another interesting FCNC process is neutral meson mixing. Since it is
an FCNC process, it cannot be mediated at tree level in the SM, and
thus it is related to the ``indirect measurements'' class of CKM
measurements. Yet, the importance of meson mixing and oscillation goes
far beyond CKM measurements and we study it in some detail.

%%%%%%%%%%%%%%%%%%%%%%%%
\subsection{Formalism}

There are four neutral mesons that can mix: $K$, $D$, $B$, and
$B_s$.\footnote{You may be wondering why there are only four meson mixing
systems. If you do not wonder and do not know the answer, then you
should wonder. We will answer this question shortly.}  We first study
the general formalism and then the interesting issues in each of the
systems. The formalism is that of a two body open system. That is, the
system involves the meson states $\Mz$ and $\Mzb$, and all the states
they can decay to. Before the meson decays the state can be a coherent
superposition of the two meson states. Once the decay happens,
coherence is practically lost. This allows us to describe the decays
using a non-Hermitian Hamiltonian, like we do for an open
system. 

We consider a general meson denoted by $P$. At $t=0$ is in an
initial state
\beq
|\psi(0)\rangle = a(0)|\Mz\rangle+b(0)|\Mzb\rangle \; ,
\eeq
where we are interested in computing the values of $a(t)$ and
$b(t)$.  Under our assumptions all the evolution is determined by a
$2\times 2$ effective Hamiltonian $\Heff$ that is not Hermitian. Any
complex matrix, such as $\Heff$, can be written in terms of Hermitian
matrices $\Meff$ and $\Geff$ as
\beq \label{defHeff}
\Heff = \Meff - \frac{i}{2}\,\Geff \; .
\eeq
$\Meff$ and $\Geff$ are associated with
$(\Mz,\Mzb)\leftrightarrow(\Mz,\Mzb)$ transitions via off-shell
(dispersive) and on-shell (absorptive) intermediate states, respectively.
Diagonal elements of $\Meff$ and $\Geff$ are associated with the
flavor-conserving transitions $\Mz\to\Mz$ and $\Mzb\to\Mzb$ while
off-diagonal elements are associated with flavor-changing transitions
$\Mz\leftrightarrow\Mzb$.

If $\Heff$ is not diagonal, the meson states, $\Mz$ and $\Mzb$ are not
mass eigenstates, and thus do not have well defined masses and
widths. It is only the eigenvectors of $\Heff$ that have well defined
masses and decay widths. We denote the light and heavy eigenstates as
$\M_L$ and $\M_H$ with $m_H > m_L$. (Another possible choice, which is
standard for $K$ mesons, is to define the mass eigenstates according
to their lifetimes: $K_S$ for the short-lived and $K_L$ for the
long-lived state. The $K_L$ is experimentally found to be the heavier
state.) Note that since $\Heff$ is not Hermitian, the eigenvectors do
not need to be orthogonal to each other. Due to CPT, $\Meff_{11} =
\Meff_{22}$ and $\Geff_{11}=\Geff_{22}$.
Then when we solve the eigenvalue problem for $\Heff$ we find that
the eigenstates are given by
\beq\label{defpq}
|\M_{L,H}\rangle=p|\Mz\rangle\pm q|\Mzb\rangle,
\eeq
with the normalization $|p|^2+|q|^2=1$ and
\beq
\left(\frac{q}{p}\right)^2=\frac{\Meff_{12}^\ast -
    (i/2)\Geff_{12}^\ast}{\Meff_{12}-(i/2)\Geff_{12}}\; .
\eeq
If CP is a symmetry of $\Heff$ then $\Meff_{12}$ and $\Geff_{12}$ are
relatively real, leading to
\beq
\left|\frac{q}{p}\right| = 1 \; ,
\eeq
where the phase of $q/p$ is unphysical. In that case the mass
eigenstates are orthogonal
\beq
\langle \M_H | \M_L\rangle = |p|^2 - |q|^2 = 0 \; .
\eeq
The real and imaginary parts of the eigenvalues of $\Heff$
corresponding to $|\M_{L,H}\rangle$ represent their masses and
decay-widths, respectively. The mass difference $\Delta m$ and the
width difference $\Delta\Gamma$ are defined as follows:
\beq\label{DelmG}
\Delta m\equiv M_H-M_L,\qquad \Delta\Gamma\equiv\Gamma_H-\Gamma_L.
\eeq
Note that here $\Delta m$ is positive by definition, while the sign of
$\Delta\Gamma$ is to be determined experimentally. (Alternatively, one
can use the states defined by their lifetimes to have
$\Delta\Gamma\equiv\Gamma_S-\Gamma_L$ positive by definition.)  The
average mass and width are given by
\beq\label{aveMG}
m\equiv{M_H+M_L\over2},\qquad \Gamma\equiv{\Gamma_H+\Gamma_L\over2}.
\eeq
It is useful to define dimensionless ratios $x$ and $y$:
\beq\label{defxy}
x\equiv{\Delta m\over\Gamma},\qquad y\equiv{\Delta\Gamma\over2\Gamma}.
\eeq
We also define
\beq \label{ee5}
\theta = \arg(M_{12} \Gamma^*_{12}).
\eeq
Solving the eigenvalue equation gives
\beq\label{eveq}
(\Delta m)^2-{1\over4}(\Delta\Gamma)^2=(4|M_{12}|^2-|\Gamma_{12}|^2),\qquad 
\Delta m\Delta\Gamma=4\re{M_{12}\Gamma_{12}^*}.
\eeq
In the limit of CP conservation, Eq.~(\ref{eveq}) is simplified to
\beq \label{derttpo}
\Delta m=2|M_{12}|, \qquad
|\Delta\Gamma|=2|\Gamma_{12}|.
\eeq

%%%%%%%%%%%%%%%%%%%%%%%%
\subsection{Time evolution}

We move on to study the time evolution of a neutral meson. For
simplicity, we assume CP conservation. Later on, when we study CP
violation, we will relax this assumption, and study the system more
generally. Many important points, however, can be understood in the
simplified case when CP is conserved and so we use it here.

In the CP limit $|q|=|p|=1/\sqrt{2}$ and we can choose the relative
phase between $p$ and $q$ to be zero. In that case the transformation
from the flavor to the mass basis, (\ref{defpq}), is simplified to
\beq
|\M_{L,H}\rangle={1 \over \sqrt{2}}\left(|\Mz\rangle\pm |\Mzb\rangle\right).
\eeq
We denote the state of an initially pure $|\Mz\rangle$ after an time
$t$ as $|\Mz(t)\rangle$ (and similarly for or $|\Mzb\rangle$). We
obtain
\beq\label{defphys-cp}
|\Mz(t)\rangle=\cos\left({\Delta E \,t\over 2}\right)|\Mz\rangle
+i \sin\left({\Delta E \,t\over 2}\right)|\Mzb\rangle\,,
\eeq
and similarly for $|\Mzb(t)\rangle$. Since flavor is not conserved,
the probability to measure a specific flavor, that is $\M$ or $\Mb$,
oscillates in time, and it is given by
\beqa \label{cp-os}
{\cal P}(\M\to \M)[t]&=&\left|\vev{\Mz(t)|\Mz}\right|^2 = 
{1+\cos(\Delta E t)\over 2},
\nonumber \\
{\cal P}(\M\to \Mb)[t]&=&\left|\vev{\Mz(t)|\Mzb}\right|^2=
{1-\cos(\Delta E t)\over 2},
\eeqa
where ${\cal P}$ denotes probability.

A few remarks are in order:
\bei
\item
In the meson rest frame, $\Delta E=\Delta m$ and  $t=\tau$, the proper time.
\item
We learn that we have flavor oscillation with frequency $\Delta m$.
This is the parameter that
eventually gives us the sensitivity to the weak interaction and to
flavor.
\item
We learn that by measuring the oscillation frequency we can determine
the mass splitting between the two mass eigenstates. One way this can be
done is by measuring the flavor of the meson both at production and
decay. It is not trivial to measure the flavor at both ends,
and we do not describe it in detail here, but you are encouraged to
think and learn about how it can be done.
\eei

%%%%%%%%%%%%%%%%%%%%%%%%
\subsection{Time scales}

Next, we spend some time understanding the different time scales that
are involved in meson mixing. One scale is the oscillation period. As
can be seen from Eq.~(\ref{cp-os}), the oscillation time scale is
given by $\Delta m$.\footnote{What we refer to here is, of course,
$1/\Delta m$. Yet, at this stage of our life as physicists, we know
how to match dimensions, and thus I interchange between time and
energy freely, counting on you to understand what I am referring to.}

Before we talk about the other time scales we have to understand how
the flavor is measured, or as we usually say it, tagged. By ``flavor
is tagged'' we refer to the decay as a flavor vs anti-flavor, for
example $b$ vs $\bar b$.  Of course, in principle, we can tag the
flavor at any time. In practice, however, the measurement is done for
us by Nature. That is, the flavor is tagged when the meson decays. In
fact, it is done only when the meson decays in a flavor specific
way. Other decays that are common to both $\M$ and $\Mb$ do not
measure the flavor. Such decays are also very useful as we will
discuss later. Semi-leptonic decays are very good flavor tags:
\beq
b \to c \mu^- \bar \nu, \qquad
\bar b \to \bar c \mu^+ \nu.
\eeq
The charge of the lepton tells us the flavor: a $\mu^+$ tells us that
we ``measured'' a $b$ flavor, while a $\mu^-$ indicates a $\bar b$. Of
course, before the meson decays it could be in a superposition of a $b$
and a $\bar b$. The decay acts as a quantum measurement. In the case
of semileptonic decay, it acts as a measurement of flavor vs
anti-flavor.

Aside from the oscillation time, one other time scale that is involved is
the time when the flavor measurements is done. Since the flavor is
tagged when the meson decays, the relevant time scale is the decay
width, $\Gamma$. We can then use the dimensionless quantity, $x\equiv
\Delta m/\Gamma$, defined in (\ref{defxy}), to understand the relevance
of these two time scales. There are three relevant regimes:
\ben
\item $x \ll 1$. We denote this case as ``slow oscillation''. In that case
the meson has no time to oscillate, and thus to good approximation
flavor is conserved. In practice, this implies that $\cos (\Delta m t)
\approx 1$ and using it in Eq. (\ref{cp-os}) we see that ${\cal
P}(\M\to \M)\approx 1$ and ${\cal P}(\M\to \Mb) \to 0$. In this case,
an upper bound on the mass difference is likely to be established
before an actual measurement. This case is relevant for the $D$
system.
%%%%%
\item $x \gg 1$.  We denote this case as ``fast oscillation''.
In this case the meson oscillates many times before decaying, and thus
the oscillating term practically averaged out to zero.\footnote{This
is the case we are very familiar with when we talk about decays into
mass eigenstates. There is never a decay into a mass eigenstate. Only
when the oscillations are very fast and the oscillatory term in the
decay rate averages out, the result seems like the decay is into a
mass eigenstate.} In practice in this case ${\cal P}(\M\to \M)\approx
{\cal P}(\M\to \Mb) \approx 1/2$ and a lower bound on $\Delta m$ can
be established before a measurement can be done. This case is relevant
for the $B_s$ system.
%%%%%%%
\item $x \sim 1$. In this case the oscillation and decay times are
roughly the same. That is, the system has time to oscillation and the
oscillation are not averaged out. In a way, this is the most
interesting case since then it is relatively easy to measure $\Delta
m$. Amazingly, this case is relevant to the $K$ and $B$ systems. We
emphasize that the physics that leads to $\Gamma$ and $\Delta m$ are
unrelated, so there is no reason to expect $x\sim 1$. Yet, Nature is
kind enough to produce $x\sim 1$ in two out of the four neutral meson
systems.
\een

It is amusing to point out that oscillations give us sensitivity to
mass differences of the order of the width, which are much smaller
than the mass itself. In fact, we have been able to measure mass
differences that are 14 orders of magnitude smaller than the
corresponding masses. It is due to the quantum mechanical nature of
the oscillation that such high precision can be achieved.

In some cases there is one more time scale: $\Delta
\Gamma$. In such cases, we have one more relevant dimensionless
parameter $y\equiv \Delta\Gamma/(2\Gamma)$. Note that $y$ is bounded,
$-1 \le y \le 1$. (This is in contrast to $x$ which is bounded by $x>0$.)
Thus, we can talk about several cases depending on the values of $y$
and $x$.
\begin{enumerate}
\item 
$|y| \ll 1$ and $y \ll x$.  In this case the width difference is
irrelevant. This is the case for the $B$ system.
\item
$y \sim x$. In this case the width different is as important as the
oscillation. This is the case in the $D$ system where $y \ll 1$ and for
the $K$ system with $y \sim 1$.
\item
$|y| \sim 1$ and $y \ll x$. In this case the oscillation averages out
and the width different shows up as a difference in the lifetime of
the two mass eigenstates. This case may be relevant to the $B_s$
system, where we expect $y \sim 0.1$.
\end{enumerate}
There are few other limits (like $y\gg x$) that are not realized in the
four meson systems. Yet, they might be realized in some other systems
yet to be discovered. 

To conclude this subsection we summarize the
experimental data on meson mixing
\beqa
x_K \sim 1, &&\qquad y_K \sim 1, \nonumber \\
x_D \sim 10^{-2}, &&\qquad y_D \sim 10^{-2}, \nonumber \\
x_d \sim 1, &&\qquad y_d \lsim 10^{-2}, \nonumber \\
x_s \sim 10, &&\qquad y_s \lsim 10^{-1}.
\eeqa
Note that $y_d$ and $y_s$ have not been measured and all we have are
upper bounds.

%%%%%%%%%%%%%%%%%%%%%%%%
\subsection{Calculation of the mixing parameters}

We now explain how the calculation of the mixing parameters is
done. We only briefly remark on $\Delta \Gamma$ and spend some time on
the calculation of $\Delta m$. As we have done a few times, 
we will do the calculation in the SM, keeping in mind that the tools
we develop can be used in a large class of models.

In order to calculate the mass and width differences, we need to know
the effective Hamiltonian, $\Heff$, defined in
Eq.~(\ref{defHeff}). For the diagonal terms, no calculations are
needed. CPT implies $M_{11}=M_{22}$ and to an excellent approximation
it is just the mass of the meson. Similarly, $\Gamma_{11}=\Gamma_{22}$
is the average width of the meson. What we need to calculate is the off
diagonal terms, that is $M_{12}$ and $\Gamma_{12}$.

We start by discussing $M_{12}$. For the sake of simplicity we
consider the $B$ meson as a concrete example.  The first point to note
is that $M_{12}$ is basically the transition amplitude between a $B$
and a $\Bb$ at zero momentum transfer. In terms of states with the
conventional normalization we have
\beq \label{mat}
M_{12} = {1 \over 2m_B} \vev{B|{\cal O}|\Bb}.
\eeq
We emphasize that we should not square the amplitude. We square
amplitudes to get transition probabilities and decay rates, which is not
the case here.

\begin{figure}
\begin{center}
\begin{picture}(210,90)(-40,-15)
\Line(0,0)(150,0)
\Line(0,50)(150,50)
%\Line(85,35)(85,-15)
%\Line(85,35)(135,35)
\Photon(50,0)(50,50) 3 5
\Photon(100,0)(100,50) 3 5
%\PhotonArc(50,80)(30,45,180) 3 5
\Text(-10,-10)[]{$b$}
\Text(160,-10)[]{$d$}
\Text(-10,60)[]{$\bar d$}
\Text(160,60)[]{$\bar b$}
\Text(73,-10)[]{$u_i$}
\Text(73,60)[]{$\bar u_j$}
%\Text(73,35)[]{$\gamma...$}
%\Text(72,5)[]{$\overline q$}
%\Text(95,120)[]{$s$}
%\Text(35,118)[]{$W$}
\end{picture}
\end{center}
\caption{A box diagram that generate an operators that can lead to $B
\leftrightarrow \Bb$ transition.\label{box}}
\end{figure}
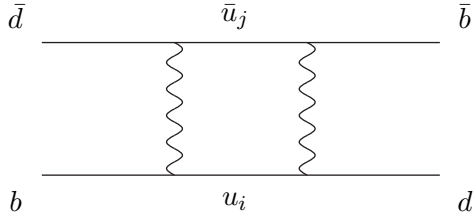

The operator that appears in (\ref{mat}) is one that can create a $B$
and annihilate a $\Bb$. Recalling that a $B$ meson is made of a $\bar
b$ and $d$ quark (and $\Bb$ from $b$ and $\bar d$), we learn that in
terms of quarks it must be of the form
\beq \label{opop}
{\cal O} \sim (\bar b \,d)(\bar b \, d).
\eeq
(We do not explicitly write the Dirac structure. Anything that does
not vanish is possible.)  Since the operator in (\ref{opop}) is an
FCNC operator, in the SM it cannot be generated at tree level and must be
generated at one loop. The one loop diagram that generates it is
called ``a box diagram'', because it looks like a square. It is
given in Fig.~\ref{box}. The calculation of the box diagram is
straightforward and we end up with
\beq \label{mat-box}
M_{12} \propto  {g^4 \over m_W^2} 
\vev{B|(\bar b_L\gamma_\mu d_L)(\bar b_L \gamma^\mu d_L)|\Bb} \,
\sum_{i,j} V_{id}^*V_{ib} V_{id}^*V_{jb} F(x_i,x_j),
\eeq
such that
\beq
x_i \equiv {m_i^2 \over m_W^2} , \qquad i=u,c,t,
\eeq
and the function $F$ is known, but we do not write it here. 

Several points are in order
\begin{enumerate}
\item
The box diagram is second order in the weak interaction, that is, it
is proportional to $g^4$.
\item
The fact that the CKM is unitary (in other words, the GIM mechanism)
makes the $m_i$ independent term vanish and to a good approximation
$\sum_{i,j}F(x_i,x_j) \to F(x_t,x_t)$. We then say that it is the top
quark that dominates the loop.
\item
The last thing we need is the hadronic matrix element,
$\vev{B|(\bar b_L\gamma_\mu d_L)(\bar b_L \gamma^\mu d_L)|\Bb}$.
The problem is that
the operator creates a free $b$ and $\bar d$ quark
and annihilates a free $\bar b$ and a $d$. This is not the same as
creating a $\Bb$ meson and annihilating a $B$ meson. Here, lattice QCD
helps and by now a good estimate of the matrix element is
available.
\item
Similar calculations can be done for the other mesons. Due to the GIM
mechanism, for the $K$
meson the function $F$ gives an extra $m_c^2/m_W^2$ suppression. 
\item
Last we mention the calculation of $\Gamma_{12}$. An estimate of it
can be made by looking at the on-shell part of the box diagram. Yet,
once particle goes on shell, QCD becomes important, and the
theoretical uncertainties in the calculation of $\Gamma_{12}$ are
larger than that of $M_{12}$.
\end{enumerate}

Putting all the pieces together we see how the measurement of the mass
difference is sensitive to some combination of CKM elements. Using the
fact that the amplitude is proportional to the heaviest internal quark
we get from (\ref{mat-box}) and (\ref{derttpo})
\beq
\Delta m_B \propto |V_{tb}V_{td}|^2,
\eeq
where the proportionality constant is known with an uncertainty at the
$10\%$ level.

%%%%%%%%%%%%%%%
\subsection{Homework}

%\vspace*{3mm}

\que{The four mesons}

It is now time to come back and ask why there are only four mesons
that we care about when discussing oscillations. In particular, why we
do not talk about oscillation for the following systems
\ben
\item
$B^+-B^-$ oscillation
\item
$K - K^*$ oscillation
\item
$T-{\ov T}$ oscillation (a $T$ is a meson made out of a $t$ and a $\bar
u$ quarks.)
\item
$K^*-{\ov K}^*$ oscillation
\een
Hint: The last three cases all have to do with time scales.  In
principle there are oscillations in these systems, but they are
irrelevant.

\vspace*{3mm}

\que{Kaons}

Here we study some properties of the kaon system. We did not talk about
it at all. You have to go back and recall (or learn) how kaons decay,
and combine that with what we discussed in the lecture.

\begin{enumerate}
\item
Explain why $y_K \approx 1$. 
\item
In a hypothetical world where we could change the mass of the kaon
without changing any other masses, how would the value of $y_K$
change if we made $m_K$ smaller or larger.
\end{enumerate}

\vspace*{3mm}

\que{Mixing beyond the SM}

Consider a model without a top quark, in which the first two
generations are as in the SM, while the left--handed
bottom ($b_L$) and the right--handed bottom ($b_R$) are $SU(2)$
singlets. 
\begin{enumerate}
\item
Draw a tree-level diagram that contributes to $B-\bar B$ mixing in this
model.
\item
Is there a tree-level diagram that contributes to  $K-\bar K$ mixing?
\item
Is there a tree-level diagram that contributes to  $D-\bar D$ mixing?
\end{enumerate}

%%%%%%%%%%%%%%%%%%%
%%%%%%% III %%%%%%%
%%%%%%%%%%%%%%%%%%%
\section{CP violation}

As we already mentioned, it turns out that in Nature CP violation is
closely related to flavor. In the SM, this is manifested by the fact
that the source of CP violation is the phase of the CKM matrix. Thus,
we will spend some time learning about CP violation in the SM and
beyond.

%%%%%%%%%%%%%%%
\subsection{How to observe CP violation?}

CP is the symmetry that relates particles with their
anti-particles. Thus, if CP is conserved, we must have
\beq \label{CP-co}
\Gamma (A \to B) = \Gamma (\bar A \to \bar B),
\eeq
such that $A$ and $B$ represent any possible initial and final states.
From this we conclude that one way to find CP violation is to
look for processes where
\beq
\Gamma (A \to B) \ne \Gamma (\bar A \to \bar B).
\eeq
This, however, is not easy. The reason is that even when CP is not
conserved, Eq.~(\ref{CP-co}) can hold to a very high accuracy in many
cases. So far there are only very few cases where (\ref{CP-co}) does not
hold to a measurable level.
The reason that it is not easy to observe CP violation is that there
are several conditions that have to be fulfilled. CP violation can
arise only in interference between two decay amplitudes. These
amplitudes must carry different weak and strong phases (we explain
below what these phases are). Also, CPT implies that the total width
of a particle and its anti-particle are the same. Thus, any CP
violation in one channel must be compensated by CP violation with an
opposite sign in another channel. Finally, it happens that in the SM,
which describes Nature very well, CP violation comes only when we have
three generations, and thus any CP violating observable must involve
all the three generations. Due to the particular hierarchical
structure of the CKM matrix, all CP violating observables are
proportional to very small CKM elements.

In order to show this we start by defining weak and strong
phases. Consider, for example, the $B \to f$ decay amplitude $A_f$,
and the CP conjugate process, $\bar B \to\fb$, with decay amplitude
$\overline{A}_{\fb}$. There are two types of phases that may appear in
these decay amplitudes.  Complex parameters in any Lagrangian term
that contributes to the amplitude will appear in complex conjugate
form in the CP-conjugate amplitude. Thus, their phases appear in $A_f$
and $\overline{A}_{\overline{f}}$ with opposite signs and these
phases are CP odd. In the SM, these phases occur only in
the couplings of the $W^\pm$ bosons and hence CP odd phases are often
called ``weak phases.''

A second type of phases can appear in decay amplitudes even when the
Lagrangian is real. They are from possible contributions of
intermediate on-shell states in the decay process. These phases are
the same in $A_f$ and $\overline{A}_{\overline{f}}$ and are therefore
CP even. One type of such phases is easy to calculate. It comes from
the trivial time evolution, $\exp(i E t)$. More complicated cases are
where there is rescattering due to the strong interactions. For this
reason these phases are called ``strong phases.''

There is one more kind of phases in additional to the weak and strong
phases discussed here. These are the spurious phases that arise due
to an arbitrary choice of phase convention, and do not originate from
any dynamics. For simplicity, we set these unphysical phases to zero
from now on.

It is useful to write each contribution $a_i$ to $A_f$ in three parts:
its magnitude $|a_i|$, its weak phase $\phi_i$, and its strong
phase $\delta_i$. If, for example, there are two such
contributions, $A_f = a_1 + a_2$, we have
\beqa\label{weastr}
A_f&=& |a_1|e^{i(\delta_1+\phi_1)}+|a_2|e^{i(\delta_2+\phi_2)},\no\\
\overline{A}_{\overline{f}}&=&
|a_1|e^{i(\delta_1-\phi_1)}+|a_2|e^{i(\delta_2-\phi_2)}.
\eeqa
Similarly, for neutral meson decays, it is useful to write
\beq\label{defmgam}
\Meff_{12} = |\Meff_{12}| e^{i\phi_M} ,\qquad
\Geff_{12} = |\Geff_{12}| e^{i\phi_\Gamma} \; .
\eeq
Each of the phases appearing in Eqs.~(\ref{weastr}) and (\ref{defmgam}) is
convention dependent, but combinations such as $\delta_1-\delta_2$,
$\phi_1-\phi_2$, and $\phi_M-\phi_\Gamma$
are physical. 
Now we can see why in order to observe CP violation we need two
different amplitudes with different weak and strong phases. It is easy
to show
%\footnote{I usually hate to use the phrase ``easy to show.''
%Too many times these points that should be easy to show can take
%weeks. Yet, in this case I really think it is easy to show. All you
%have to do is find the magnitude of two complex number and show when
%they are not equal.}  
and I leave it for the homework.

A few remarks are in order:
\ben
\item
The basic idea in CP violation research is to find processes where we
can measure CP violation. That is, we look for processes with two
decay amplitudes that are roughly of the same size with different
weak and strong phases.
\item
In some cases, we can get around QCD. In such cases, we get sensitivity
to the phases of the unitarity triangle (or, equivalently, of the CKM
matrix). These cases are the most interesting ones.
\item
Some observables are sensitive to CP phases without measuring CP
violation. That is like saying that we can determine the angles of a
triangle just by knowing the lengths of its sides.
\item
While we talk only about CP violation in meson oscillations and decays,
there are more types of CP violating observables. In particular,
triple products and electric dipole moments (EDMs) of elementary
particles encode CP violation. They are not directly related to
flavor, and are not covered here.
\item
So far CP violation has been observed only in meson decays,
particularly, in $K_L$, $B_d$ and $B^\pm$ decays. In the following, we
concentrate on the formalism relevant to these systems.
\een

%%%%%%%%%%%%%%%%%%%%%%%%
\subsection{The three types of CP violation}

When we consider CP violation in meson decays there are two types of
amplitudes: mixing and decay. Thus, there must be three ways to
observe CP violation, depending on which type of amplitudes
interfere. Indeed, this is the case. We first introduce the three
classes and then discuss each of them in some length.
\ben
\item
CP violation in decay, also called direct CP violation. This
is the case when the interference is between two decay amplitudes. The
necessary strong phase is due to rescattering. 
\item
CP violation in mixing, also called indirect CP violation. In this
case the absorptive and dispersive mixing amplitudes interfere. The
strong phase is due to the time evolution of the oscillation.
\item
CP violation in interference between mixing and decay. As the name
suggests, here the interference is between the decay and the
oscillation amplitudes. The dominant effect is due to the dispersive
mixing amplitude (the one that gives the mass difference) and a
leading decay amplitude. Also here the strong phase is due to the time
evolution of the oscillation.
\een

In all of the above cases the weak phase comes from the Lagrangian. In
the SM these weak phases are related to the CKM phase. In many cases,
the weak phase is one of the angles of the unitary
triangle.

%%%%%%%%%%%%%%%%%%%%%
\subsection{CP violation in decay}

We first talk about CP violation in decay.  This is the case when
\beq
|A(\M \to f)| \ne |A(\Mb \to \fb)|. 
\eeq
The way to measure this type of CP violation is as follows. We define
\beq
a_{CP}\equiv{\Gamma( \bar B \to \bar f)-\Gamma(B\to f) \over
\Gamma( \bar B \to \bar f)+\Gamma(B\to f)}={|\bar A/A|^2-1\over|\bar A/A|^2+1}.
\eeq
Using  (\ref{weastr}) with $\phi$ as the weak phase difference and
$\delta$ as the strong phase difference, we write
\beq
A(\M\to f)=A\left(1+r \exp[i(\phi+\delta)]\right), \qquad
A(\M\to \fb)=A\left(1+r \exp[i(-\phi+\delta)]\right),
\eeq
with $r\le 1$.
We get
\beq\label{acp-dec}
a_{CP}=r \sin\phi\sin\delta.
\eeq
This result shows explicitly that we need two decay amplitudes, that is, $r \ne
0$, with different weak phases, $\phi\ne0,\pi$ and different strong phases
$\delta\ne 0,\pi$.

A few remarks are in order:
\begin{enumerate}
\item
In order to have a large effect we need each of the three
factors in (\ref{acp-dec}) to be large. 
\item
CP violation in decay can occur in both charged and neutral
mesons. One complication for the case of neutral meson is that it is
not always possible to tell the flavor of the decaying meson, that is,
if it is $\M$ or $\Mb$. This can be a problem or a virtue.
\item
In general the strong phase in not calculable since it is related to
QCD. This may not be a problem if all we are after is to demonstrate
CP violation. In other cases the phase can be independently measured,
eliminating this particular source of theoretical error.
\end{enumerate}

\subsubsection{$B \to K \pi$}
\begin{figure*}[t]
\begin{center}
\begin{picture}(510,165)(0,-20)
\Line(0,80)(50,80)
\Line(50,80)(87,117)
\Line(85,35)(85,-15)
\Line(85,35)(134,35)
\Gluon(50,80)(85,35) 3 5
\PhotonArc(50,80)(30,45,180) 3 5
\Text(10,70)[]{$b$}
\Text(57,52)[]{$g$}
\Text(72,5)[]{$\overline q$}
\Text(85,99)[]{$s$}
\Text(120,47)[]{$q$}
\Text(40,140)[]{($P$)}
\Line(165,80)(215,80)
\Line(215,80)(252,117)
\Line(250,35)(250,-15)
\Line(250,35)(300,35)
\Photon(215,80)(250,35) 3 5
\Text(175,70)[]{$b$}
\Text(222,52)[]{$W$}
\Text(237,5)[]{$\overline u$}
\Text(250,98)[]{$u$}
\Text(285,47)[]{$s$}
\Text(205,140)[]{($T$)}
\Line(330,80)(380,80)
\Line(380,80)(417,117)
\Line(415,35)(415,-15)
\Line(415,35)(464,35)
\Photon(380,80)(415,35) 3 5
\PhotonArc(380,80)(30,45,180) 3 5
\Text(340,70)[]{$b$}
\Text(385,52)[]{$Z,\gamma$}
\Text(402,5)[]{$q$}
\Text(415,99)[]{$s$}
\Text(450,47)[]{$q$}
\Text(360,140)[]{($P_{EW}$)}
\end{picture}
\end{center}
\caption[y]{The $B \to K \pi$ amplitudes. The dominant one is
the strong penguin amplitude ($P$), and the sub-dominant ones are the
tree amplitude ($T$) and the electroweak penguin amplitude ($P_{EW}$).
\label{bkpi-fig}}
\vspace{-0.45cm}
\end{figure*}
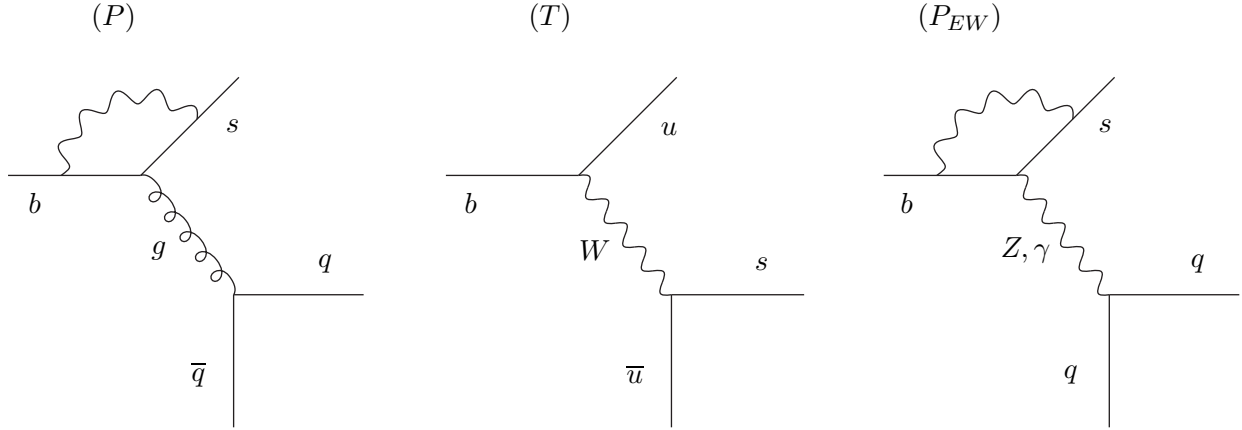

Our first example of CP violation in decay is $B^0 \to K^+ \pi^-$. At
the quark level the decay is mediated by $b \to s \bar u u$
transition. There are two dominant decay amplitudes, tree level and
one loop penguin diagrams.\footnote{This is the first time we introduce
the name penguin. It is just a name, and it refers to one loop
amplitude of the form $f_1 \to f_2 B$ where $B$ is a neutral boson
that can be on--shell or off--shell. If
the boson is a gluon we may call it QCD penguin. When it is a photon
or a $Z$ boson it is called electroweak penguin.} Two penguin diagrams
and the tree level diagram are plotted in fig.~\ref{bkpi-fig}.

Naively, tree diagrams are expected to dominate. Yet, this is not the
case here. The reason is that the tree diagram is highly CKM
suppressed. It turns out that this suppression is stronger than the
loop suppression such that $r=|P/T|\sim 0.3$. (Here we use $P$ and $T$
to denote the penguin and tree amplitudes.) In terms of weak phases,
the tree amplitude carries the phase of $V_{ub} V_{us}^*$. The
dominant internal quark in the penguin diagram is the top quark and
thus to first approximation the phase of the penguin diagram is the
phase of $V_{tb} V_{ts}^*$, and to first approximation $\phi=\alpha$.
As for the strong phase, we cannot calculate it, and there is no
reason for it to vanish since the two amplitudes have different
structure.  Experimentally, CP violation in $B \to K \pi$ decays has
been established. It was the first observation of CP violation in
decay.

We remark that $B \to K \pi$ decays have much more to offer. There are
four different such decays, and they are all related by isospin, and
thus many predictions can be made. Moreover, the decay rates s are
relatively large and the measurements have been performed. The
full details are beyond the scope of these lectures, but you are
encouraged to go and study them.

%%%%%%%%%%%%%%%%%%%%%
\subsubsection{$B \to D K$}

Our second example is $B \to D K$ decay. This decay involves only tree
level diagrams, and is sensitive to the phase between the $b \to c
\bar u s$ and $ b \to u \bar c s$ decay amplitude, which is
$\gamma$. The situation here is involved as the $D$ further
decays and what is measured is $B \to f_D K$, where $f_D$ is a final
state that comes from a $D$ or $\ov D$ decay. This ``complication''
turns out to be very important. It allows us to construct
theoretically very clean observables. In fact, $B \to D K$ decays are
arguably the cleanest measurement of a CP violation phase in terms of
theoretical uncertainties.

The reason for this theoretical cleanliness is that all the necessary
hadronic quantities can be extracted experimentally. We consider
decays of the type
\beq
B \to D (\ov D)\, K (X) \to f_D\, K (X),
\eeq
where $f_D$ is a final state that can be accessible from both $D$ and
$\ov D$ and $X$ represents possible extra particles in the final
state. The crucial point is that in the intermediate state the flavor
is not measured. That is, the state is in general a coherent
superposition of $D$ and $\ov D$. On the other hand, this state is
on-shell so that the $B\to D$ and $D\to f_D$ amplitudes
factorize. Thus, we have quantum coherence and factorization at the
same time. The coherence makes it possible to have interference and
thus sensitivity to CP violating phases. Factorization is important
since then we can separate the decay chain into stages such that each
stage can be determined experimentally. The combination is then very
powerful, we have a way to probe CP violation without the need to
calculate any decay amplitude.

To see the power of the method, consider using $B \to DKX$ decays with
$n$ different $X$ states, and $D \to f_D$ with $k$ different $f_D$
states, one can perform $n\times k$ measurements. Because the $B$ and
$D$ decay amplitude factorize, these $n \times k$ measurements depend
on $n+k$ hadronic decay amplitudes. For large enough $n$ and $k$,
there is a sufficient number of measurements to determine all hadronic
parameters, as well as the weak phase we are after. Since all hadronic
matrix elements can be measured, the theoretical uncertainties are
much below the sensitivity of any foreseeable future experiment.

%%%%%%%%%%%%%%%%%%%%%
\subsection{CP violation that involves mixing}

We move on to study CP violation that involves mixing. This kind of CP
violation is the one that was first discovered in the kaon system in
the 1960s, and in the $B$ system more recently. They are the ones that
shape our understanding of the picture of CP violation in the SM, and
thus, they deserve some discussion.

We start by re-deriving the oscillation formalism in a more general
case where CP violation is included. Then we will be able to construct
some CP violating observables and see how they are related to the
phases of the unitarity triangle. For simplicity we concentrate on
the $B$ system. We allow the decay to be into an arbitrary state,
that is, a state that can come from any mixture of $B$ and
$\Bb$. Consider a final state $f$ such that
\beq
A_f \equiv A(B \to f), \qquad
\bar A_f \equiv A(\Bb \to f).
\eeq
We further define 
%$\tau$ to be the proper time and
\beq\label{deflam}
\lambda_f \equiv \frac{q}{p}\frac{\overline{A}_f}{A_f} \; .
\eeq
We consider the general time evolution of a $\Mz$ and $\Mzb$
mesons. It is given by
\beqa\label{defphys}
|\Mz(t)\rangle&=&g_+(t)\,|\Mz\rangle
- (q/p)\ g_-(t)|\Mzb\rangle,\no\\
|\Mzb(t)\rangle&=&g_+(t)\,|\Mzb\rangle
- (p/q)\ g_-(t)|\Mz\rangle \; ,
\eeqa
where we work in the $B$ rest frame and
\beq
g_\pm(t) \equiv \frac12\left(e^{-im_Ht-\frac12\Gamma_Ht}\pm
  e^{-im_Lt-\frac12\Gamma_Lt}\right).
\eeq
We define $\tau \equiv \Gamma t$ and then the decay rates are
\beqa \label{cprate}
\Gamma(B \to f)[t] = |A_f|^2 e^{-\tau}&& \hspace*{-3mm}{\Big\{}
(\cosh y\tau + \cos x\tau) + |\lambda_f|^2 (\cosh y\tau - \cos x\tau) 
\nonumber \\ && \hspace*{-2mm}-
2 {\Re}\left[\lambda_f (\sinh y\tau + i\sin x\tau)\right]{\Big\}},
\nonumber \\
\Gamma(\bar B \to f)[t] = |\bar A_f|^2 e^{-\tau}&& \hspace*{-3mm}{\Big\{}
(\cosh y\tau + \cos x\tau) + |\lambda_f|^{-2} (\cosh y\tau - \cos x\tau) 
\nonumber \\ && \hspace*{-2mm}-
2 {\Re}\left[\lambda_f^{-1} (\sinh y\tau + i\sin x\tau)\right]{\Big\}},
\eeqa
where $\Gamma(B \to f)[t]$ ($\Gamma(\Bb \to f)[t]$) is the
probability for an initially pure $B$ ($\Bb$) meson to decay at time
$t$ to a final state $f$.

%From that we get the following time-dependent decay rates:
%\beqa
%\frac{\Gamma[\Mz(t)\to f]}{e^{-\Gamma t}}&=&
%\left(|A_f|^2+|(q/p)\overline{A}_f|^2\right)\cosh(y\Gamma t)
%  +\left(|A_f|^2-|(q/p)\overline{A}_f|^2\right)\cos(x\Gamma t)\no\\
%&+&2\,\re{(q/p)A_f^\ast \overline{A}_f}\sinh(y\Gamma t)
%-2\,\im{(q/p)A_f^\ast \overline{A}_f}\sin(x\Gamma t)
%\label{decratbt1}\;,\\
%\frac{\Gamma[\Mzb(t)\to f]}{e^{-\Gamma t}}&=&
%\left(|(p/q)A_f|^2+|\overline{A}_f|^2\right)\cosh(y\Gamma t)
%  -\left(|(p/q)A_f|^2-|\overline{A}_f|^2\right)\cos(x\Gamma t)\no\\
%&+&2\,\re{(p/q)A_f\overline{A}^\ast_f}\sinh(y\Gamma t)
%-2\,\im{(p/q)A_f\overline{A}^\ast_f}\sin(x\Gamma t)
%\label{decratbt2}\; ,
%\eeqa
Terms proportional to $|A_f|^2$ or $|\overline{A}_f|^2 $ are
associated with decays that occur without any net oscillation, while
terms proportional to $|\lambda|^2$ or $|\lambda|^{-2}$ are associated
with decays following a net oscillation. The $\sinh(y\tau)$ and
$\sin(x\tau)$ terms in Eqs.~(\ref{cprate}) are associated with the
interference between these two cases. Note that, in multi-body decays,
amplitudes are functions of phase-space variables. The amount of
interference is in general a function of the kinematics, and can be
strongly influenced by resonant substructure.  Eqs.~(\ref{cprate}) are
much simplified in the case where $|q/p|=1$ and
$|A_f/\overline{A}_f|=1$.  In that case $|\lambda|=1$ is a pure phase.

We define the CP observable of asymmetry of neutral meson
decays into final CP eigenstates $f$
\beq\label{asyfcp}
{\cal A}_{f}(t)\equiv\frac{\Gamma[\Bb(t)\to f]- 
\Gamma[B(t)\to f]}
{ \Gamma[\Bb(t)\to f]+ \Gamma[B(t)\to f]}.
\eeq
If $\Delta\Gamma = 0$ and $|q/p|=1$, as expected to a good
approximation for $B$ system, and the decay amplitudes
fulfill $|\overline{A}_{f}|=|A_{f}|$, the interference
between decays with and without mixing is the only source of the
asymmetry and
\beq \label{cpin}
{\cal A}_{f}(t)=\im{\lambda_f}\sin(x\tau) =
\sin\left[\arg(\lambda_f)\right]\sin(\Delta m t).
\eeq
where in the last step we used $|\lambda|=1$. We see that once we
know $\Delta m$, and if the above conditions are satisfied, we have a
clean measurement of the phase of $\lambda_f$. This phase is directly
related to an angle in the unitarity triangle, as we discuss shortly.

It is instructive to describe the effect of CP violation in decays of
the mass eigenstates. For cases where the width difference is
negligible, this is usually not very useful. It is not easy to
generate, for example, a $B_H$ mass eigenstate. When the width
difference is large, like in the kaon system, this representation can
be very useful as we do know how to generate $K_L$ states. We assume
$|q/p|=1$ and then the decay into CP eigenstates is given by
\beq \label{massdecay}
\Gamma(\M_{L} \to f_{\CP}) = 2 |A_f|^2 e^{-\Gamma_{L}t} \cos^2\theta, \qquad
\Gamma(\M_{H} \to f_{\CP}) = 2 |A_f|^2 e^{-\Gamma_{H}t} \sin^2\theta.
\eeq
where 
\beq
\theta \equiv {\arg{\lambda} \over 2}.
\eeq
In this case, CP violation, that is $\theta \ne 0$, is manifested by
the fact that two non-degenerate states can decay to the same CP
eigenstate final state.

We also consider decays into a pure flavor state. In that
case $\lambda=0$, and we can isolate the effect of CP violation in
mixing. CP violation in mixing is defined by
\beq\label{cpvmix}
|q/p|\neq1 \; .
\eeq
This is the only source of CP violation in charged-current 
semileptonic neutral meson decays $\M,\Mb\to \ell^{\pm} X$. 
This is because we use 
$|A_{\ell^+ X}|=|\overline{A}_{\ell^- X}|$ and $A_{\ell^- X} =
\overline{A}_{\ell^+ X} = 0$, which, to lowest order in $G_F$, is the
case in the SM and in most of its extensions, and thus
$\lambda=0$.

This source of CP violation can be measured via the asymmetry of
``wrong-sign'' decays induced by oscillations:
\beq\label{asysl}
{\cal A}_{\rm SL}(t)\equiv\frac{\Gamma[
\Bb(t)\to\ell^+X]-\Gamma[B(t)\to\ell^-X]}
{\Gamma[\Bb(t)\to\ell^+X]+\Gamma[B(t)\to\ell^-X]}
=\frac{1-|q/p|^4}{1+|q/p|^4}.
\eeq
Note that this asymmetry of time-dependent decay rates is actually
time independent. 

We are now going to give few examples of cases that are sensitive to
CP violation that involve mixing.

%%%%%%%%%%%%%%%%%%%%%%%%%%%%%%%%%%%%
\subsubsection{$B \to \psi K_S$}

The ``golden mode'' with regard to CP violation in interference
between mixing and decays is $B \to \psi K_S$. It provides a very clean
determination of the angle $\beta$ of the unitarity triangle.

As we already mentioned we know that to a very good approximation
in the $B$ system $|q/p|=1$. In that case we have
\beq\label{lamhad}
\lambda_f=e^{-i\phi_B}{\overline{A}_f\over A_f} \; ,
\eeq
where $\phi_B$ refers to the phase of $M_{12}$ [see
Eq.~(\ref{defmgam})]. Within the SM, where the top diagram dominates
the mixing, the corresponding phase factor is given to a very good
approximation by
\beq\label{phimsm}
e^{-i\phi_B}={V_{tb}^* V_{td}^{}\over V_{tb}^{}V_{td}^*} \;.
\eeq
For $f=\psi K$, which proceeds via a $\bar b\to \bar cc\bar s$
transition, we can write 
\beq\label{btoccs}
A_{\psi K}=\left(V^\ast_{cb} V^{}_{cs}\right)T_{\psi
  K}
\eeq
where $T_{\psi K}$ is the magnitude of the tree amplitude. In
principle there is also a penguin amplitude that contributes to the
decay. The leading penguin carries the same weak phase as the tree
amplitude. The one that carries a different weak phase is highly CKM
suppressed and we neglect it. This is a crucial point. Because of that
we have to a very good approximation
\beq
|\lambda_{\psi K_S}|=1, \qquad \Im(\lambda_{\psi K_S})=\sin2\beta,
\eeq
and Eq.~(\ref{cpin}) can be used for this case. We conclude that a
measurement of the CP asymmetry in $B \to \psi K_S$ gives a very clean
determination of the angle $\beta$ of the unitarity triangle. Here we
were able to overcome QCD by the fact that the decay is dominated by
one decay amplitude that cancels once the CP asymmetry is
constructed. As for the strong phase it arises due to the oscillation
and it is related to the known $\Delta m$. This CP asymmetry
measurement was done and is at present the most precise measurement of
any angle or side of the unitarity triangle.

A subtlety arises in this decay that is related to the fact that
${B}^0$ decays into $\psi K^0$ while $\overline{B}^0$ decays into
$J/\psi\overline{K}{}^0$. A common final state, e.g. $J/\psi K_S$, is
reached only via $K^0-\overline{K}{}^0$ mixing. We do not elaborate on
this point.

There are many more decay modes where a clean measurement of angles
can be performed. In your homework, you will work out one more example
and even try to get an idea of the theoretical errors.

%%%%%%%%%%%%%%%%%%%%%%%%%%%
\subsubsection{$K$ decays}

CP violation was discovered in $K$ decays, and until recently, it was
the only meson where CP violation had been measured. CP violation
were first observed in $K_L\to\pi\pi$ decays in 1964, and later in
semileptonic $K_L$ decays in 1967. Beside the historical importance,
kaon CP violation provides important bounds on the unitarity
triangle. Moreover, when we consider generic new physics, CP violation
in kaon decays provides the strongest bound on the scale of the new
physics. This is a rather interesting result based on the amount of
progress that has been made in our understanding of flavor and CP
violation in the last 45 years.

While the formalism of CP violation is the same for all mesons, the
relevant approximations are different. For the $B$ system, we neglected
the width difference and got the very elegant formula,
Eq.~(\ref{cpin}). For the $B$ mesons it is easy to talk in terms of
flavor (or CP) eigenstates, and use mass eigenstates only as
intermediate states to calculate the time evolutions.
For kaons, however, the width difference is very large 
\beq
{\Gamma_S \over \Gamma_L} \sim 600.
\eeq
This implies that, to very good approximation, we can get a state that
is pure $K_L$. All we have to do is wait. Since we do have a pure
$K_L$ state, it is easy to talk in terms of mass eigenstates. Note
that it is not easy to get a pure $K_S$ state. At short times we have
a mixture of states and, only after the $K_S$ part has decayed, we have
a pure $K_L$ state.

In terms of mass eigenstates, CP violation is manifested if the same
state can decay to both CP even and CP odd states. This should be
clear to you from basic quantum mechanics. Consider a symmetry, that
is, an operator that commutes with the Hamiltonian. In the case under
consideration, if CP is a good symmetry it implies $[\mbox
{CP},H]=0$. When this is the case, any non-degenerate state must be an
eigenstate of CP. In a CP-conserving theory, any eigenstate of CP must
decay to a state with the same CP parity. In particular, it is
impossible to observe a state that can decay to both CP even and CP
odd states. Thus, CP violation in kaon decays was established when
$K_L \to \pi \pi$ was observed. $K_L$ decays dominantly to three pions,
which is a CP odd state. The fact that it decays also to two pions,
which is a CP even state, implies CP violation.

We do not get into the details of the calculations, but it must be
clear at this stage that the rate of $K_S \to \pi \pi$ must be related
to the values of the CKM parameters and, in particular, to its
phase. I hope you will find the time to read about it in one of the
reviews I mentioned.

Before concluding, we remark on semileptonic CP violation in
kaons. When working with mass eigenstates, CP conservation implies that
\beq
\Gamma(K_L \to \pi^- \ell^+ \nu)=
\Gamma(K_L \to \pi^+ \ell^- \bar\nu).
\eeq
(If it is not clear to you why CP implies the above, stop for a second
and convince yourself.) Experimentally, the above equality was found
to be violated, implying CP violation.

In principle, the CP violation in $K_L \to \pi\pi$ and in semileptonic
decay are independent observables. Yet, when all the decay amplitudes
carry the same phase, these two are related. This is indeed the case
in the kaon system, and thus we talk about one parameter that measure
kaon CP violation, which is denoted by $\varepsilon_K$. (Well, there
is one more parameter called $\varepsilon'_K$, but we will not discuss
it here.)

%%%%%%%%%%%%%%%%%%%%%%%%%%%
\subsection{Homework}

\que{Condition for CP violation}

Using Eq.~(\ref{weastr}), show that in order to observe CP violation,
$\Gamma(B \to f) \ne \Gamma(\Bb \to \fb)$, we need two amplitudes with
different weak and strong phases.

\vspace*{3mm}

\que{Mixing formalism}

%\underline{Question 1:} 
\vspace*{2mm}

In this question, you are asked to develop the general formalism of
meson mixing.

\begin{enumerate}
\item
Show that the mass and width differences are given by
\beq \label{ee3}
4(\Delta m)^2 - (\Delta \Gamma)^2 = 4 (4|M_{12}|^2 - |\Gamma_{12}|^2), \qquad
\Delta m \Delta \Gamma = 4 \Re(M_{12}\Gamma_{12}^*),
\eeq
and that 
\beq \label{ee4}
\left|{q \over p}\right| = 
\left|{\Delta m - i \Delta \Gamma/2} \over 2M_{12} - i \Gamma_{12}\right| .
\eeq
\Answer{I define
\beq
H=\pcmatrix{a & b \cr c & a}
\eeq
From the eigenvalue equation we get
\beq
(a-\mu)^2-bc=0 \ \gorer \ \mu = a \pm \sqrt{bc}\ 
\gorer \ \Delta \mu = 2 \sqrt{bc}
\eeq
From one of the the eigenfunction equations we get
\beq
(a-\mu)p+bq=0 \ \gorer \ {q \over p} = \pm {\Delta \mu \over 2 b}
\eeq
Which gives (\ref{ee4}).
Using 
\beq
4bc = (2M_{12} - i \Gamma_{12}) 
(2M_{12}^* - i \Gamma_{12}^*)
= 4|M_{12}|^2- |\Gamma_{12}|^2 - 4 i \Re(M_{12} \Gamma^*_{12})
\eeq
and
\beq
(\Delta \mu)^2 = (\Delta m)^2 + {(\Delta \Gamma)^2 \over 4} 
- i (\Delta m \Delta \Gamma)
\eeq
Then from $(\Delta \mu)^2 = 4 b c$ and comparing the real and imaginary parts
we get (\ref{ee3}).}
\item
When CP is a good symmetry all  mass eigenstates must 
also be CP eigenstates.
Show that CP invariance requires 
\beq
\left|{q \over p}\right| = 1.
\eeq
\Answer{under CP $\ket{B} \to \ket{\bar B}$  (up to an unphysical phase
which I set to zero).
Thus,
\beq
\ket{B_1} = p \ket{B^0} + q\ket{\bar B^0} \ \underCP \
q \ket{B^0} + p\ket{\bar B^0}
\eeq
and it is clear that in order for $\ket{B_1}$ to stay invariant under
CP transformation we need $|q|=|p|$.}
\item
In the limit $\Gamma_{12} \ll M_{12}$ show that 
\beq
\Delta m = 2 |M_{12}|, \qquad
\Delta \Gamma = 2 |\Gamma_{12}| \cos \theta, 
\qquad
\left|{q \over p}\right| = 1.
% +O\left(\Gamma_{12} \over M_{12}\right)
\eeq
\Answer{
From (\ref{ee3}) we can solve for $\Delta m$ and $\Delta \Gamma$ and we get
\beq
(\Delta \Gamma)^2=
2\left(\sqrt{a^2+b^2}-a\right), \qquad
(\Delta m)^2=
{1 \over 2}\left(\sqrt{a^2+b^2}+a\right),
\eeq
where
\beq
a=4|M_{12}|^2- |\Gamma_{12}|^2, \qquad b=4 \Re(M_{12} \Gamma^*_{12})
\eeq
It is now clear how to take the required limit.}
\item 
Derive Eqs.~(\ref{cprate}).
\item
Derive Eq.~(\ref{cpin}).
\item 
Show that when $\Delta \Gamma = 0$ and $|q/p|=1$
\beqa
\Gamma(B \to X \ell^- \bar \nu)[t]&=& 
e^{-\Gamma t} \sin^2 (\Delta m t/2), \nonumber \\
\Gamma(B \to X \ell^+ \nu)[t] &=& e^{-\Gamma t} \cos^2 (\Delta m t/2).
\eeqa
\Answer{
In the  $|q/p|=1$ case we have
\beq
\ket{B} = {1 \over \sqrt{2}}\left(\ket{B_1} +\ket{B_2}\right) , \qquad
\ket{\bar B} = {1 \over \sqrt{2}}\left(\ket{B_1} -\ket{B_2}\right) .
\eeq
The mass eigenstate evolve according to 
\beq
a_i(t)=a_i(0) \exp[(\gamma/2 + i M_i)t]
\eeq
Since $a_i(0)=1/\sqrt{2}$ we find
\beqa
P(B \to \bar B) &=& |a_1(t)-a_2(t)|^2 = e^{-\Gamma t} \sin^2 (\Delta m t/2),
\nonumber \\
P(B \to B) &=& |a_1(t)+a_2(t)|^2 = e^{-\Gamma t} \cos^2 (\Delta m t/2) 
\eeqa}
\end{enumerate}

\vspace*{3mm}

\que{$B \to \pi^+\pi^-$ and CP violation}

One of the interesting decays to consider is $B \to \pi\pi$. Here we
only briefly discuss it.
\begin{enumerate}
\item
First assume that there is only tree level decay amplitude (that is,
neglect penguin amplitudes).  Draw the Feynman diagram of the
amplitude, paying special attention to its CKM dependence.
\item
In that case, which angle of the unitarity 
triangle is the time dependent CP asymmetry, Eq.~(\ref{cpin}), 
sensitive to?
\item
Can you estimate the error introduced by neglecting the penguin
amplitude?
(Note that one can use isospin to reduce this error. Again, you are
encouraged to read about it in one of the reviews.)
\end{enumerate}

\vspace*{3mm}

\que{$B$ decays and CP violation}

Consider the decays $\bar B^0 \to \psi K_S$ and $\bar B^0 \to \phi
K_S$. Unless explicitly noted, we always work within the framework of
the standard model. 
%Recall the following quark assignment for the
%relevant hadrons:
%\beq
%\bar B^0(b \bar d), \qquad
%\psi (c \bar c), \qquad
%\phi (s \bar s), \qquad
%K_S={1 \over \sqrt{2}}(K + \bar K), \qquad
%K (d \bar s), \qquad
%\bar K (s \bar d).
%\eeq
%
\begin{enumerate}
\item
$\bar B^0 \to \psi K_S$ is a tree-level process. Write down the underlying
quark decay. Draw the tree level diagram. What is the CKM dependence
of this diagram? In the Wolfenstein parametrization, what is the weak
phase of this diagram?
\item
Write down the underlying quark decay for $B^0 \to \phi K_S$. Explain
why there is no tree level diagram for $B^0 \to \phi K_S$.
\item
The leading one loop diagram for $B^0 \to \phi K_S$ is a gluonic
penguin diagram. As we have discussed, there are several diagrams and
only their sum is finite. Draw a representative diagram with an
internal top quark. What is the CKM dependence of the diagram? In the
Wolfenstein parametrization, what is the weak phase of the diagram?
\item
Next we consider the time dependent CP asymmetries. We define as usual
\beq
\lambda_f \equiv {\bar A_f \over A_f} {q\over p}, \qquad
A_f \equiv A(B^0\to f) ,\qquad
\bar A_f \equiv A(\bar B^0\to f).
\eeq
In our case we neglect subleading diagrams and then we have
$|\lambda|=1$ and thus
\beq
a_f\equiv
{\Gamma(\bar B^0(t) \to f) -\Gamma( B^0(t) \to f) \over
  \Gamma(\bar B^0(t) \to f) +\Gamma( B^0(t) \to f)}  =
- \Im\lambda_f \,\sin(\Delta m_B\, t)
\eeq
Both $a_{\psi K_S}$ and $a_{\phi K_S}$ measure the same angle of
the unitarity triangle. That is, in both cases, $\Im\lambda_f = \sin 2
x$ where $x$ is one of the angles of the unitarity triangle. What is
$x$? Explain.
\item
Experimentally,
\beq
\Im\lambda_{\psi K_S}= 0.68(3), \qquad
\Im\lambda_{\phi K_S}= 0.47(19).
\eeq
Comment about these two results. In particular,
do you think these two results are in disagreement?
\item
Assume that in the future we will find
\beq
\Im\lambda_{\psi K_S}= 0.68(1), \qquad
\Im\lambda_{\phi K_S}= 0.32(3).
\eeq
That is, that the two results are not the same.
Below are three possible ``solutions''. For each solution explain if
you think it could work or not. If you think it can work, show how. If
you think it cannot, explain why.
\begin{enumerate}
\item
There are standard model corrections that we neglected.
\item
There is a new contribution to $B^0 -\bar B^0$ mixing with a weak phase that
is different from the SM one.
\item
There is a new contribution to the gluonic penguin with a weak phase that
is different from the SM one.
\end{enumerate}
%\Answer{
%\begin{enumerate}
%\item
%In the SM the correction are versy small. The small parameter
%is $|V_{ub}V_{us}|V_{cb}V_{cs}| < 10^{-2}|$, so it is very unlikely to
%make significant change of the naive prediction.
%\item
%New contribution to $B -\bar B$ mixing would affect both asymmetries
%the same, so in that case we will still have the equal.
%\item
%New contribution to the penguin
%decay would affect only $a_{\phi K_S}$ and could cause the
%difference.
%\end{enumerate}
%}
\end{enumerate}

\que{Decay of mass eigenstates}

Derive Eq.~(\ref{massdecay}). The idea is to understand that when we
talk about mass eigenstates, we are talking about ``late times,'' $t
\gg x\Gamma$ so that the $\sin(\Delta m t)$ term can be averaged out.

%%%%%%%%%%%%%%%%%%%%%%%%%%%
\section{Putting it all together: The big picture}

After we got a taste of how to probe the flavor sector, we are ready
to ask: what are the results? That is, how compatible are all the
measurements? As we explained, in principle we need four
measurements to define the flavor parameters and the rest are checks
on the model. Then we can ask what are the implications of these
results on the big question of physics, that is, what is the fundamental
Lagrangian of Nature.

%%%%%%%%%%%%%%%%%%%%%%%%%%%
\subsection{The current status of the SM flavor sector}

Out of the four flavor parameters of the SM, two are known to high
accuracy, and there is a very good agreement between the various ways they
are determined. These two parameters are $\lambda$ and $A$ of the
Wolfenstein parametrization of the CKM matrix. Thus, it is customary
to plot all the other measurements as bounds on the rescaled unitarity
triangle, that depend only on two parameters, $\rho$ and
$\eta$. There are many measurements and bounds and within the errors
each of them gives an allowed range in the $\rho-\eta$ plane for the
one undetermined apex of the rescaled unitarity triangle.

What I am about to show you (in the next page, please do not peek!) is the
most recent compilation of all these results. I am sure you have 
seen it before. Yet, I hope that now you appreciate it much more. In
class, I always stop at this point. I like to make sure the students
do not miss the big moment, and that they see how amazing physics is. If I
knew how to play a trumpet, this would be the moment that I would use
it. In a written version, it is a bit harder. I cannot stop you from
looking at the next page and I certainly cannot play the trumpet here.
Still I do ask you to take a break here. Make sure you fully understand
what you are going to see.

\begin{figure}[t]
\centerline{\includegraphics[width=11.9cm]{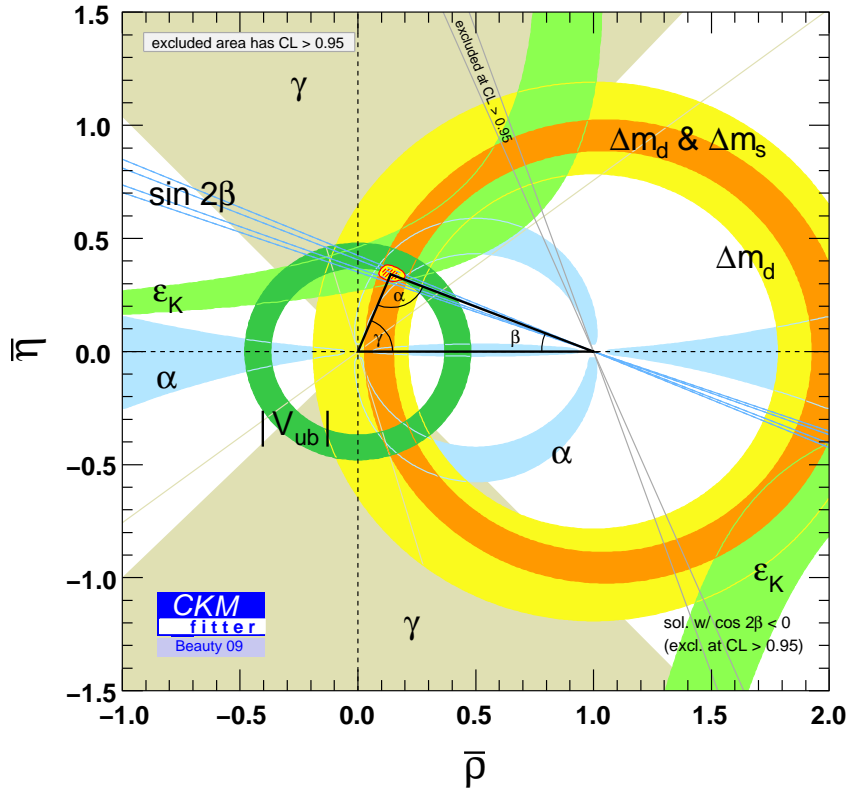}}
\caption[y]{Global fit to the unitarity triangle based on all
available data. (Taking from the CKMfitter group
website, ckmfitter.in2p3.fr.)
\label{global-fit-fig}}
\end{figure}

Now, that you are ready, take your time and look at
Fig.~\ref{global-fit-fig}. What you see
are many bounds all overlapping at one small area in the $\rho-\eta$
plane. That is, taking the two measurements as determining $\rho$ and
$\eta$, all the rest are checks on the SM. You see that the flavor
sector of the SM passes all its tests. Basically, all the measurements
agree.

The most important implication of this triumph of theoretical and
experimental physics is the following statement: The
Cabibbo-Kobayashi-Maskawa mechanism is the dominant source of flavor
and CP violation in low-energy flavor-changing processes.  This is a
very important statement. Indeed the Nobel prize was awarded to
Kobayashi and Maskawa in 2008 because it is now experimentally proven
that the KM phase is the one which explains the observed CP violation in
Nature.

%%%%%%%%%%%%%%%%%%%%%%

\subsection{Instead of a summary: The NP flavor problem}

The success of the SM can be seen as a proof that it is an effective
low energy description of Nature. There are, however, many reasons to
suspect that the SM has to be extended. A partial list includes the
hierarchy problem, the strong CP problem, baryogenesis, gauge coupling
unification, the flavor puzzle, neutrino masses, and gravity. We are
therefore interested in probing the more fundamental theory. One way
to go is to search for new particles that can be produced at yet
unreached energies. Another way to look for new physics is to search
for indirect effects of heavy unknown particles. Flavor physics is
used to probe such indirect signals of physics beyond the SM.

In general, flavor bounds provide strong constraints on new physics
models.  This fact is called ``the new physics flavor problem''.  The
problem is actually the mismatch between the new physics scale that is
required in order to solve the hierarchy problem and the one that is
needed in order to satisfy the experimental bounds from flavor
physics.

In order to understand what the new physics flavor problem is, let us
first recall the hierarchy problem. In order to prevent the Higgs mass
from getting a large radiative correction, new physics must appear at
a scale that is a loop factor above the weak scale
\beq \label{hpscale}
\Lambda \lsim 4 \pi m_W \sim 1 \; {\rm TeV}. 
\eeq
Here, and in what follows, $\Lambda$ represents the new physics scale.
Note that such TeV new physics can be directly probed in collider searches.

While the SM scalar sector is unnatural, its flavor sector is
impressively successful.\footnote{The flavor structure of the SM is
interesting since the quark masses and mixing angles exhibit
hierarchy.  These hierarchies are not explained within the SM, and
this fact is usually called ``the SM flavor puzzle''. This puzzle is
different from the new physics flavor problem that we are discussing
here.} This success is linked to the fact that the SM flavor structure
is special. As we already mentioned, the charged current interactions
are universal (in the mass basis, this is manifest through the
unitarity of the CKM matrix) and FCNCs are highly suppressed:
they are absent at the tree level and at the one loop level they are
further suppressed by the GIM mechanism.  These special features are
important in order to explain the observed pattern of weak decays.
Thus, any extension of the SM must conserve these successful features.

Consider a generic new physics model, where the only suppression of
FCNCs processes is due to the large masses of the particles that
mediate them. Naturally, these masses are of the order of the new
physics scale, $\Lambda$. Flavor physics, in particular measurements
of meson mixing and CP violation, puts severe constraints on $\Lambda$.
In order to find these bounds we take an effective field theory
approach. At the weak scale we write all the non-renormalizable
operators that are consistent with the gauge symmetry of the SM. In
particular, flavor-changing four Fermi operators of the form (the
Dirac structure is suppressed)
\beq \label{genfour}
{q_1 \bar q_2 q_3 \bar q_4 \over  \Lambda^2},
\eeq
are allowed. Here $q_i$ can be any quark as long as the
electric charges of the four fields in Eq. (\ref{genfour}) sum up to
zero. We emphasize that there is no exact symmetry that can
forbid such operators. This is in contrast to operators that violate
baryon or lepton number that can be eliminated by imposing symmetries
like $U(1)_{B-L}$ or R-parity.
The strongest bounds are obtained from meson mixing and CP-violation
measurements. Depending on the mode we find bounds of the order
\beq\label{fla-lam}
\Lambda \gtrsim \mbox{few}\;\times10^{4} {\mbox{~TeV}}.
\eeq
There is tension between the new physics scale that is required in
order to solve the hierarchy problem, Eq. (\ref{hpscale}), and the one
that is needed in order not to contradict the flavor bounds,
Eq.~(\ref{fla-lam}).  The hierarchy problem can be solved with new
physics at a scale $\Lambda \sim 1$ TeV.  Flavor bounds, on the other
hand, require $\Lambda \gtrsim 10^4$ TeV.  This tension implies that any TeV
scale new physics cannot have a generic flavor structure. This is the
new physics flavor problem.

Flavor physics has been mainly an input to model building,
not an output. The flavor predictions of any new physics model are 
not a consequence of its generic structure but rather of the special 
structure that is imposed to satisfy the severe existing flavor bounds.
It is clearly a very interesting open question to determine the NP
model and how it deals with flavor.

\subsection{Concluding remarks}
This is a good time to finish the lectures. I hope that you gained some
understanding of flavor physics, how it was used to shape the SM as we
know it, and why it is so important in our quest to find the theory that
extends the SM. In the near future we expect more data in the energy
frontier, as well as more flavor data. It can be really fun to see how
the two can work together to show us what Nature is at really short
distances, that is, to help us in getting a better answer to the
fundamental question of physics
\beq
{\cal L}=?
\eeq

%%%%%%%%%%%%
\section*{Acknowledgements}
I thank Joshua Berger, Mario Martone, and Dean Robinson for comments
on the manuscript. The work of YG is supported by NSF grant number
PHY-0757868 and by a U.S.-Israeli BSF grant.

%%%%%%%%%%%%%%%%%%%%%

\end{document}